\documentclass[11pt]{article}
\usepackage{graphicx}
\usepackage{amssymb}
\usepackage{amsmath}
\usepackage[margin=1in]{geometry}
\usepackage{float}
\usepackage{bm}
\usepackage{commath}
\usepackage{xcolor}
\usepackage{authblk}
\title{Interaction between deformation twinning and dislocation slip in polycrystalline solids}

\author{Eric Ocegueda}

\author{Kaushik Bhattacharya}

\affil{Division of Engineering and Applied Science,\\
	California Institute of Technology, \\Pasadena, CA 91125}

\begin{document}
\maketitle

\begin{abstract}
Deformation twinning is a form of permanent deformation that is commonly observed in low symmetry crystals such as hexagonal close-packed (hcp) metals. With recent increased interest in using hcp metals, such as magnesium, in structural, automotive, and armor applications due to their high strength to weight ratio, there is a need for a comprehensive understanding of deformation twinning and its interaction with dislocation slip. A great deal has been learned at the microscopic level where individual dislocations interact with twin boundaries through atomistic simulations, and at the macroscopic level by ignoring morphology and treating twinning as `pseudo-slip'. However, twins form collectively across multiple grains with complex morphology that affects the bulk behavior. These mesoscale aspects have been less studied and are the focus of this paper. We present a model that describes the twin and slip morphology, its evolution, and interactions in a unified manner at the scale of several grains and use it to study the implications on macroscopic behavior. The key ideas are to combine a phase-field model of twinning with a crystal plasticity model of slip, and to implement it in parallel on graphic processing units for fast computations.

\end{abstract}


\section{Introduction}\label{sec:Introduction}

The recent decades have seen an interest in alloys with low-symmetry, more specifically hexagonal close-packed crystal structure, for structural applications due to their high strength-to-weight ratio. For example, magnesium alloys have amongst the highest strength to weight ratio (with a density of 1.8 g/cm$^3$ and yield strength exceeding 100 MPa) of known metals and have been explored for automotive, biomedical, and other engineering applications. However, these alloys often have limited ductility and suffer sudden, almost brittle, failure. We refer the reader to recent reviews \cite{Joost2017, Kulekci2008a, Kusnierczyk2017, Chen2016}.

The high strength to weight ratio as well as limited ductility has its origins in the anisotropy of the low symmetry crystals and the resulting complexity of deformation modes (e.g. \cite{Rissanen2012}). For example, magnesium, which is hexagonal close-packed, has an easy basal slip system, relatively easy tension twins but hard pyramidal and prismatic systems. Unfortunately, the basal slip and tension twin systems are deficient, i.e., they can not accommodate an arbitrary deviatoric strain. Therefore, a polycrystal of these materials needs to engage the hard slip systems resulting in high strength. However, the mismatch between the strength of these systems leads to strain localization and other related phenomena that cause easy failure. This is in contrast with face-centered cubic crystals where the basal slip systems are complete.

Twinning adds additional complexity. Unlike slip, which involves the sliding of one plane of atoms over the other and is carried by dislocations or line defects, twins are planar defects across which a shear restores the lattice. We refer the reader to Mahajan and Christian \cite{Christian1995} for a comprehensive introduction to twinning. Thus twins manifest themselves as bands, in contrast, to slip that is typically more diffuse, leading to length-scale effects that are distinct from slip. Further, twins involve a rotation of the lattice. Finally, the shear in twinning has a specific sense (i.e., it can shear in one direction but not the other), while slip does not. Therefore, twinning can lead to an asymmetric response to imposed loading. The complexity is compounded by the interaction between slip and twinning, especially in polycrystal domains. Therefore, the interaction between twinning and slip in low symmetry crystals has been a subject of interest in recent years.

One line of work has focused on the atomistic scale where the energetics of twinning, nucleation of twins, the structure of twin boundaries, and the interaction of individual dislocations with a twin boundary are determined \cite{Li2009, Cao2006, Jiang2021, Yamakov2002, Tang2020}. While these studies provide important insights and inputs to larger-scale models, they are insufficient to describe deformation morphology and overall macroscopic response.

There is also a large body of work at the polycrystalline scale. Since Kalidindi \cite{Kalidindi1998}, various researchers \cite{Agnew2001, Staroselsky2003,Graff2007, Zhang2012, Chang2015, Feng2018, Feng2020} have studied the interaction between slip and twinning by treating twinning as a ``pseudo-slip''. Briefly, pseudo-slip does not seek to describe the morphology of individual twins but only an average twin volume fraction, allowing the incorporation of twins into full-field crystal plasticity or self-consistent studies of polycrystalline behavior. These studies have been invaluable in understanding the relationship between texture and the strength of the material. However, a detailed comparative study by Abdolvand and Daymond \cite{abdolvand2013} on twinning in Zircaloy-2 between electron back-scatter diffraction (EBSD) observations and pseudo slip based crystal plasticity simulations showed significant differences in the evolution of the twins.  Further, these models they do not describe the morphology which plays a critical role in material failure.
The role of morphology, especially through the cooperative actions of grains, has been emphasized by several researchers \cite{Beyerlein2011,Orozco-Caballero2017}.

Phase-field models have long been used to study morphology in the context of phase transitions (e.g. \cite{ajk_actamat_01}). However, this has largely been in the context of small (geometrically linear) strains, which are not appropriate for deformation twins that can involve large shears and lattice rotations. Therefore, the phase-field approaches have also been extended to finite deformation in the context of twinning (e.g. \cite{Clayton2011}) and separately slip-based plasticity (e.g. \cite{Beyerlein2016}). Recently, Liu {\it et al.} \cite{Liu2018} proposed a phase-field model that combines deformation twinning and plastic slip, and used it to study both single crystals in three dimensions and polycrystals in two dimensions. They emphasized the role of grain boundaries in twin nucleation and transmission into neighboring poorly oriented grains. However, their model still uses an incremental update for the twinning deformation borrowing from the pseudo-slip models. An alternate approach, by Jin {\it et al.} \cite{Jin2019} inserted fully twinned regions (via strain discontinuities) when appropriate stochastic nucleation and propagation conditions are met. Two dimensional polycrystal simulations correctly captured twin nucleation and propagation from grain boundaries under dynamic loading conditions. However, this model does not fully capture the nucleation and growth of twins and thus misses the full interaction with dislocations.

In this work, we develop a phase-field model to investigate the interactions between twinning and slip at the scale of multiple grains. We follow Mahajan and Christian \cite{Christian1995} to describe twinning deformation and combine it with plastic slip. We incorporate general energetic and kinetic laws to describe nucleation barriers, surface energy, propagation drag, and rate hardening of twins and slip. We propose an implementation that is massively parallel and allows the use of graphical processing units (GPUs) to conduct large-scale studies. We then use the model and its implementation to conduct detailed studies in two dimensions that provide insights into the interaction between the two deformation mechanisms. The focus of this work is understanding the interaction and the resulting consequences on polycrystalline media.
\section{Model} \label{sec:PhaseFieldModel}
We first describe the model for a single crystal and then extend it to polycrystals.

\subsection{Kinematics}
Consider a single crystal undergoing a deformation $\mathbf{y}$ from a stress-free reference configuration. We assume that the deformation gradient ($\mathbf{F} = \nabla\mathbf{y}$) is multiplicatively decomposed into elastic, plastic (related to slip), and twinning parts,

\begin{equation}
	\mathbf{F} = \mathbf{F}^{\, \text{e}}\mathbf{F}^{\, \text{p}}\mathbf{F}^{\, \text{t}}.
\end{equation}
This decomposition naturally handles the situation where slip follows twinning at any material point and matches the common decomposition within literature \cite{Feng2018, Feng2020, Jin2019}. The other scenario, where twinning follows slip, requires the incorporation of transmutations and is ignored in this work, and will be addressed in future work.

\begin{figure}
	\centering
	\includegraphics[width=0.425\linewidth]{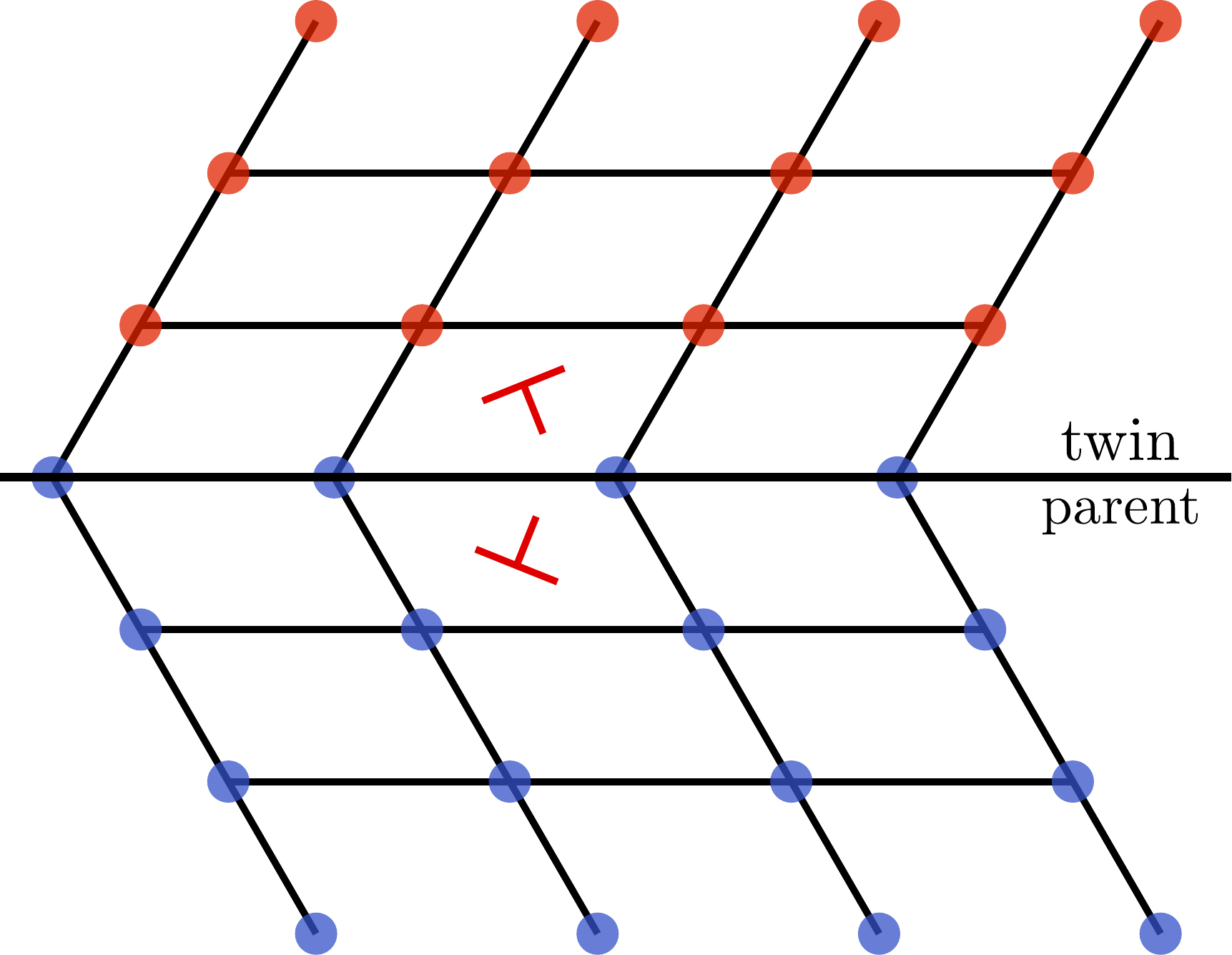}\\
	\caption{Twinning schematic. Parent and twinned lattice are shown with the respective dislocation slips in red. }
	\label{fig:TwinningSchematic}
\end{figure}

We assume that we have one twin system and one slip system. The {\it twin system} is characterized by a twinning shear magnitude $\gamma^{\, \text{t}}_0$, twinning shear direction $\mathbf{\hat{b}}^{\, \text{t}}$, twin plane normal $\mathbf{\hat{n}}^{\, \text{t}}$, and twinning rotation $\mathbf{R}$. In this work, we consider a type I twin so $\mathbf{R} = 2\hat{\mathbf{n}}^{\, \text{t}}\otimes\hat{\mathbf{n}}^{\, \text{t}} - \mathbf{I}$ is a two-fold rotation about the twin plane. The twinning deformation is
\begin{equation}
\mathbf{F}^{\, \text{t}}(\eta) = \mathbf{I} + \eta \, \gamma^{\, \text{t}}_0 \, \mathbf{\hat{b}}^{\, \text{t}}\otimes\mathbf{\hat{n}}^{\, \text{t}},
\end{equation}
where $\eta=0$ in the untwinned region and $\eta=1$ in the twinned region. We assume that $\eta$ is smooth and changes from 0 to 1 in a narrow region with a corresponding interpolation of the twinning deformation. The {\it slip system} is characterized by the slip activity $\gamma^{\, \text{p}}$, slip direction $\mathbf{\hat{b}}^{\, \text{p}}$ (respectively $\mathbf{\hat{R}} \mathbf{\hat{b}}^{\, \text{p}}$), and glide plane normal $\mathbf{\hat{n}}^{\, \text{p}}$ (respectively $\mathbf{\hat{R}} \mathbf{\hat{n}}^{\, \text{p}}$) in the untwinned (respectively twinned) region, see Figure~\ref{fig:TwinningSchematic}. We describe $\mathbf{\hat{b}}^{\, \text{p}}$ and $\mathbf{\hat{n}}^{\, \text{p}}$ by the angle, $\theta^p$, $\mathbf{\hat{b}}^{\, \text{p}}$ makes to a reference and use the interpolation 
\begin{equation}
	\theta(\eta) = \eta\theta(1) + (1 - \eta)\theta(0)
\end{equation}
to obtain an unified description. The plastic deformation evolves according to
\begin{equation}
\dot{\mathbf{F}}^{\, \text{p}} = \big(\dot{\gamma}^{\, \text{p}}\mathbf{\hat{b}}^{\, \text{p}} (\eta) \otimes \mathbf{\hat{n}}^{\, \text{p}} (\eta) \big)\mathbf{F}^{\, \text{p}}.
\end{equation}
Finally, we introduce the accumulated plastic activity $\epsilon^{\, \text{p}}$ which evolves according to $\dot{\epsilon}^{\, \text{p}} = |\dot{\gamma}^{\, \text{p}}|$. Note that $\eta \ge 0$ since twinning shear has a specific sense while $\gamma^{\, \text{p}}$ can be both negative and positive since slip can lead to shear in either sense.

\subsection{Free energy and stress}
The free energy density is postulated to be 
\begin{equation}
	W( \mathbf{F}^{\, \text{e}},  \eta, \nabla\eta, \epsilon^{\, \text{p}}) = W_{\, \text{e}}(\mathbf{F}^{\, \text{e}}, \eta) + \frac{\alpha}{2}\left\lVert \nabla\eta\right\rVert^2 +  W_{\, \text{t}}(\eta) + W_{\, \text{p}}(\epsilon^{\, \text{p}}),
\end{equation}
where
\begin{equation}
    W_{\, \text{e}}(\mathbf{F}^{\, \text{e}},\eta) = \frac{1}{2}\mathbf{E}^{\, \text{e}}:{\mathbb{C}}(\eta):\mathbf{E}^{\, \text{e}}
\end{equation}
is the elastic energy density with non-linear strain measure $\mathbf{E}^{\, \text{e}} = ((\mathbf{F}^{\, \text{e}})^\intercal \mathbf{F}^{\, \text{e}} - \mathbf{I})/2$ and (phase-dependent anisotropic) elastic modulus ${\mathbb{C}}(\eta)$; 
\begin{equation}
	W_{\, \text{t}}(\eta) = \frac{M}{2}\eta^2(\eta - 1)^2  \label{eq:TwinningEnergy}
\end{equation}
is a double-well potential governing the phase-field variable \cite{Clayton2011, Yang2010}, and
\begin{equation}
W_{\, \text{p}}(\epsilon^{\, \text{p}}) = \sigma^{\, \text{p}}_{\infty} \bigg[ \epsilon^{\, \text{p}} + \frac{\sigma^{\, \text{p}}_{\infty}}{h} \exp\Big( {-h\frac{\epsilon^{\, \text{p}}}{\sigma^{\, \text{p}}_\infty}}\Big) \bigg]
\end{equation}
describes the plastic hardening similar to the Voce hardening law \cite{Agnew2001, Graff2007}.

The Piola-Kirchhoff stress is given by 
\begin{equation}
\mathbf P = \frac{\partial W^{\, \text{e}}}{\partial {\mathbf{F}}} = \frac{\partial W^{\, \text{e}}}{\partial {\mathbf{F}^{\, \text{e}}}} ({\mathbf F}^{\, \text{p}}{\mathbf F}^{\, \text{t}})^{-\intercal}.
\end{equation}

\subsection{Equilibrium and evolution}
Mechanical equilibrium requires
\begin{equation} \label{eq:equil}
\nabla \cdot {\mathbf P} =  {\mathbf 0}.
\end{equation}
The evolution of the internal variables $\gamma^{\, \text{p}}$ and $\eta$ follow the evolution equations (flow rule and kinetic relation)
\begin{align}
    0 &\in - \tau^{\, \text{p}}+\frac{\partial W_{\, \text{p}}}{\partial \epsilon^{\, \text{p}}} 
     + \partial_{\dot{\epsilon}^{\, \text{p}}} \Psi^{*}  \label{eq:SlipStationary}\\
    0 &\in -\tau^{\, \text{t}} + \frac{\partial W_{\, \text{e}}}{\partial \eta} + \frac{\partial W_{\, \text{t}}}{\partial \eta} - \alpha \nabla^2 \eta
    + \partial_{\dot{\eta}} \Psi^{*},  \label{eq:TwinStationary}
\end{align}
where $\tau^{\, \text{p}}$ and $\tau^{\, \text{t}}$ are the plastic and twin resolved shear stress respectively defined as
\begin{align}
    &\tau^{\, \text{p}} = \bm{\sigma}^{\, \text{p}} : (\mathbf{\hat{b}}^{\, \text{p}} \otimes \mathbf{\hat{n}}^{\, \text{p}}) , \;\;\; \bm{\sigma}^{\, \text{p}} = (\mathbf{F}^{\, \text{e}})^{\,\intercal}\mathbf{P}\big(\mathbf{F}^{\, \text{p}}\mathbf{F}^{\, \text{t}}\big)^\intercal,\\
    &\tau^{\, \text{t}} = \bm{\sigma}^{\, \text{t}} : (\mathbf{\hat{b}}^{\, \text{t}} \otimes \mathbf{\hat{n}}^{\, \text{t}}) , \;\;\; \bm{\sigma}^{\, \text{t}} = \gamma^{\, \text{t}}_0\big( \mathbf{F}^{\, \text{e}}\mathbf{F}^{\, \text{p}}\big)^\intercal \mathbf{P}.
\end{align}
$ \Psi^*$ is the dissipation potential we postulate to be
\begin{equation}
    \Psi^*\big(\dot{\gamma}^{\, \text{p}}, \dot{\eta} \big) = \bigg[ \tau^{\, \text{p}}_0|\dot{\gamma}^{\, \text{p}}| + \frac{\tau^{\, \text{p}}_0\dot{\gamma}^{\, \text{p}}_0}{m_{\, \text{p}} + 1} \bigg(\frac{|\dot{\gamma^{\, \text{p}}}|}{\dot{\gamma}^{\, \text{p}}_0}\bigg)^{m_{\, \text{p}} + 1} \bigg] + \; \bigg[ \tau^{\, \text{t}}_0|\dot{\eta}| + \frac{\tau^{\, \text{t}}_0\dot{\gamma}^{\, \text{t}}_0}{m_{\, \text{t}} + 1} \bigg(\frac{|\dot{\eta}|}{\dot{\gamma}^{\, \text{t}}_0}\bigg)^{m_{\, \text{t}} + 1}\bigg] , \label{eq:Psi*}
\end{equation}
$\tau^{\, \text{p}}_0$ and $\tau^{\, \text{t}}_0$ are the critical resolved shear stresses, $\dot{\gamma}^{\, \text{p}}_0$ and $\dot{\gamma}^{\, \text{t}}_0$ the reference shear rate, and $m_{\, \text{p}}$ and $m_{\, \text{t}}$ are power rate hardening parameters for slip and twinning respectively.

Note that the dissipation potential $ \Psi^*$ is not continuously differentiable when $\dot{\gamma}^{\, \text{p}}=0$ or $\dot{\eta}=0$ and therefore (\ref{eq:SlipStationary}), (\ref{eq:TwinStationary}) are formulated as differential inclusions.  In particular, it means that there is no evolution of the plastic strain unless the resolved shear stress exceeds the critical values.

\subsection{Remarks on twinning}\label{sec:TwinNucleation}
The double-well energy, $W_{\, t}$, and gradient energy follow the Allen-Cahn model \cite{Allen1979} and lead to a twin boundary whose thickness is on the order of $\sqrt{\alpha/M}$ ($\sim 8\sqrt{\alpha/M}$ in our numerical studies -- see supplementary information) and whose energy density is on the order of $\sqrt{\alpha M}$ per unit area.

The twin evolution equation (\ref{eq:TwinStationary}) becomes the equilibrium equation of \cite{Clayton2011} if we chose $\Psi^*=\Psi^*(\dot{\gamma}^{\, \text{p}})$ and the time-dependent Landau Ginzburg equation of \cite{Miranville2003} if we chose $\Psi^*= \Psi^*_{\, \text{p}}(\dot{\gamma}^{\, \text{p}}) + \nu/2 |\dot{\eta}|^2$. We chose the form (\ref{eq:Psi*}) because it provides both a critical stress $\tau^{\, \text{t}}_0$ for the propagation of an existing twin boundary and a critical stress of $\tau^{\, \text{t}}_0 + M/(6\sqrt{3})$ for twin nucleation. To see the latter, consider a material in the parent (untwinned) state with $\eta=0$ uniformly and subjected to extremely slow loading. According to (\ref{eq:TwinStationary}), $\eta$ will evolve only when the resolved shear stress matches
\begin{equation}
    \tau^{\, \text{t}} =  M\eta(\eta - 1)(2\eta - 1) + \tau^{\, \text{t}}_0.
    \label{eq:TwinNucleation}
\end{equation}
Thus, the nucleation of a twin, $\eta \rightarrow 1$, requires the resolved shear stress to overcome the maximum value of the first term (which is $M/(6\sqrt{3})$) in addition to $\tau^{\, \text{t}}_0$. We have confirmed this numerically.

In short, our model of twinning provides a twin boundary energy $\sqrt{\alpha M}$ per unit area, critical stress for nucleation $\tau^{\, \text{t}}_0 +M/(6\sqrt{3})$, and a critical stress for propagation $\tau^{\, \text{t}}_0$.

\subsection{Polycrystal domains}

The polycrystal is an assemblage of grains made of the same material but whose orientation differs from each other. We describe the texture of the polycrystal using the orientation function ${\mathbf Q}({\mathbf X})$ that is the rotation that takes the grain at ${\mathbf X}$ to a fiducial grain. The slip and twin systems in the grain ${\mathbf X}$ are now described by ${\mathbf Q}({\mathbf X}) \mathbf{\hat{b}}^{\, \text{p}}, {\mathbf Q}({\mathbf X}) \mathbf{\hat{n}}^{\, \text{p}}, {\mathbf Q}({\mathbf X}) \mathbf{\hat{b}}^{\, \text{t}},{\mathbf Q}({\mathbf X}) \mathbf{\hat{n}}^{\, \text{t}}$ and the free energy density $W( \mathbf{F}^{\, \text{e}},  \eta, \nabla\eta, \epsilon^{\, \text{p}}, {\mathbf X})  = W( \mathbf{F}^{\, \text{e}} {\mathbf Q} ({\mathbf X}),   \eta, \nabla\eta {\mathbf Q}({\mathbf X}), \epsilon^{\, \text{p}}) $.

\subsection{Numerical implementation}
We consider a periodic domain and prescribe a time-dependent average deformation gradient $\bar{\mathbf F}(t)$. So, ${\mathbf y} = \tilde{\mathbf y} + \bar{\mathbf F}{\mathbf X}$ where $\tilde{\mathbf y}$ is periodic, as are the rest of the kinematic quantities. We follow the accelerated computational micromechanics approach of Zhou and Bhattacharya \cite{Zhou2020} to cast the governing equations as a time discretized variational problem, which is solved using fast Fourier transforms and implemented on graphical processing units (GPUs). When solving the deformation and internal variable evolution at each time step, the previous configuration is augmented with a small perturbation of order $10^{-4}$ to assist in convergence. We provide additional details in the Appendix.

\section{Results} \label{sec:MicrostructureEvolCubic}

\begin{table}
\begin{center}
	\begin{tabular}{ |p{2cm}||p{4cm}|p{5cm}|p{3cm}| }
		\hline
		Parameter & Value & Significance & Reference\\
		\hline
		\multicolumn{4}{|c|}{Elastic Parameters} \\
		\hline
		$\lambda_1$ & $25$ GPa & Stiffness $C_{1111}$ term & \cite{Tutcuoglu2019} \\
		$\lambda_2$ & $15$ GPa & Stiffness $C_{1122}$ term & \cite{Tutcuoglu2019} \\
		$\mu$ & $15$ GPa & Stiffness $C_{1212}$ term & \cite{Tutcuoglu2019} \\
		\hline
		\multicolumn{4}{|c|}{Twinning Parameters} \\
		\hline
		$M$ & $80$ MPa & Double well & -- \\
		$\alpha$ & $1.25\times10^{-8}$ GPa$\cdot$ nm$^2$ & Surface energy & \cite{Wang2009, Levitas2009, Kronberg1968}\\
		$\gamma^{\, \text{t}}_0$ & $0.129$ & Shear magnitude & \cite{Tutcuoglu2019}\\
		$\dot{\gamma}_0^{\, \text{t}}$ & $1.0$ $1/s$ & Reference shear rate & \cite{Tutcuoglu2019} \\
		$m_{\, \text{t}}$ & $1.0$ & Rate hardening & \cite{Tutcuoglu2019}\\
		$\tau^{\, \text{t}}_0$ & $1$ MPa & Critical resolved shear stress & \cite{Chang2017}\\
		$\theta^{\, \text{t}}$ & $-\frac{\pi}{8}$ & Twin shear angle & --\\
		\hline
		\multicolumn{4}{|c|}{Plasticity Parameters} \\
		\hline
		$m_{\, \text{p}}$ & $0.05$ & Rate hardening & \cite{Chang2015}\\
		$\dot{\gamma}_0^{\, \text{p}}$ & $1.0$ $1/s$ & Reference shear rate & \cite{Chang2015}\\
		$\tau^{\, \text{p}}_0$ & $4$ MPa & Critical resolved shear stress & \cite{Chang2015}\\
		$\sigma^{\, \infty}$ & $2$ MPa & Ultimate slip stress & \cite{Chang2015}\\
		$h$ & $7.1$ GPa & Hardening constant & \cite{Chang2015}\\
		$\theta^{\, \text{p}}$ & $\frac{\pi}{8}$ & Slip shear angle & --\\
		\hline
	\end{tabular}\\
	\caption{Material parameters used for the simulations unless otherwise specified.}
	\label{tab:PhaseFieldParameters}
\end{center}
\end{table}

We now present results that highlight various aspects of microstructure evolution and, in particular, the interplay between slip and twinning in a model two dimensional polycrystalline system. Material parameters are given in Table \ref{tab:PhaseFieldParameters}, unless otherwise specified, and are motivated by basal slip and tension twinning in magnesium. Calculations are conducted on a $1024 \times 1024$ grid that offers sufficient resolution to resolve twin boundaries. Various simulations validating and verifying the model and implementation are presented in supplementary materials.

\subsection{Typical results}\label{sec:typical}
\begin{figure}
	\centering
    \includegraphics[width=\linewidth]{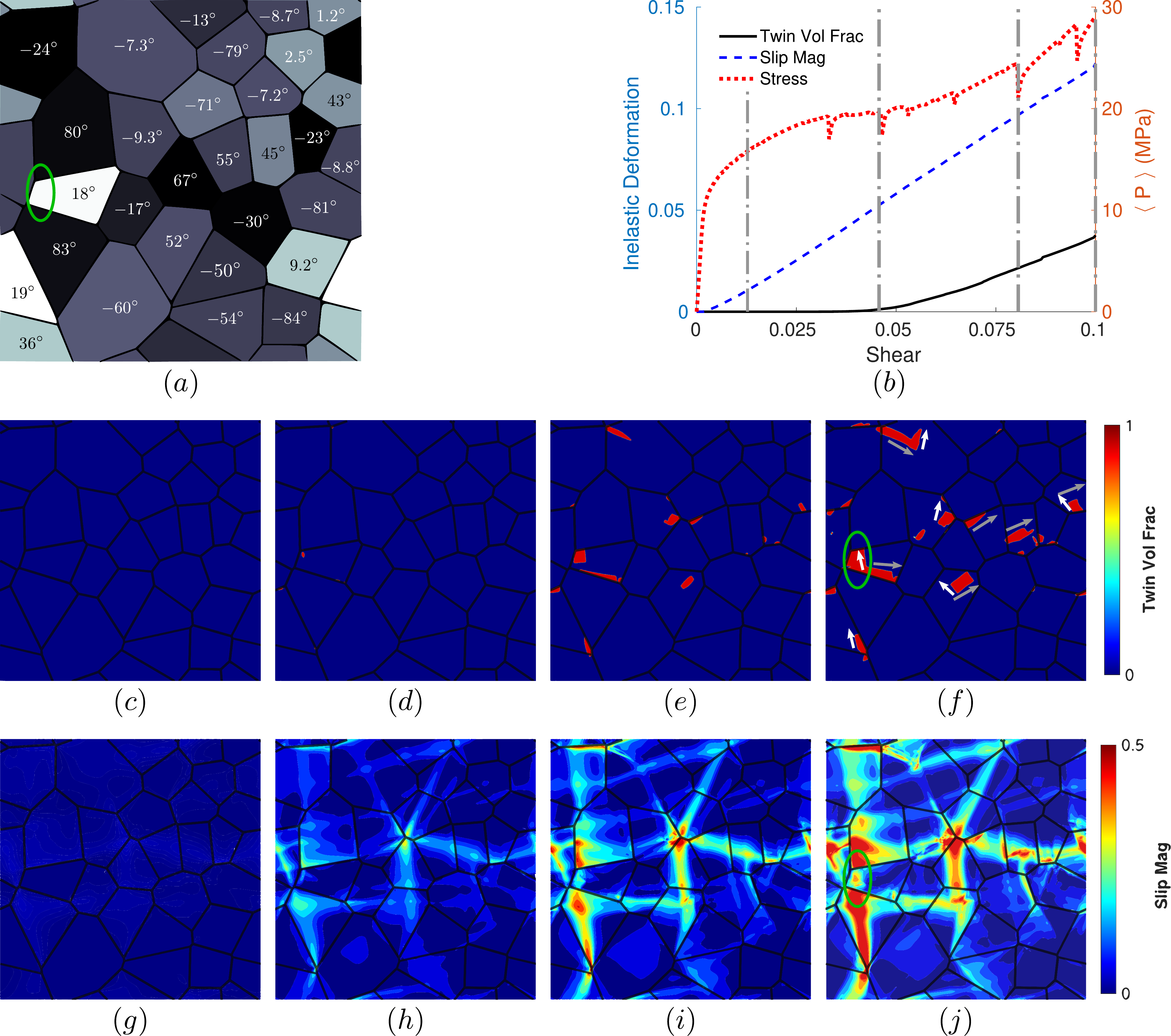}
	\caption{Deformation of polycrystalline specimen subjected to simple shear. (a) The polycrystalline specimen. The grayscale shows the propensity for twinning and slip. The darker grains are favorably oriented for slip and lighter grains for twinning, while the numbers indicate the grain orientation. (b) Stress-strain behavior and evolution of average slip magnitude and twin volume fraction. Vertical dashed lines indicate points at which microstructure is shown. (c--f) Snapshot of twin volume fraction ($\eta$) at strains of $\gamma = 0.01, \, 0.044, \, 0.07, \, 0.1$. Gray and white arrows show twin and reciprocal twin directions respectively. (g--j) Snapshots of slip magnitude ($\gamma^{\, \text{p}}$) at strains of $\gamma = 0.01, \, 0.044, \, 0.07, \, 0.1$.}
	\label{fig:PhaseFieldMicrostructure}
\end{figure}

We consider a polycrystal of 30 grains, shown in Figure~\ref{fig:PhaseFieldMicrostructure}(a), obtained by Voronoi tessellation from random seeds and sample the orientations from a uniform distribution between $\pm\frac{\pi}{2}$. We have verified that the results are typical for this texture by considering other samples -- see supplementary information. We apply a boundary condition corresponding to a simple shear along the horizontal axis at a strain rate of $1.0 \times 10^{-2}$ until a maximum shear strain of $0.1$. The stress-strain behavior, bulk slip activity, and twin volume fraction are shown in Figure~\ref{fig:PhaseFieldMicrostructure}(b). Snapshots of the spatial evolution of twinning and slip are shown in Figure~\ref{fig:PhaseFieldMicrostructure}(c--f) and \ref{fig:PhaseFieldMicrostructure}(g--j) respectively.

As loading begins, the response is initially elastic till the applied strain reaches a value of about $\gamma = 0.01$. At this point, plastic slip initiates across several grains -- Figure~\ref{fig:PhaseFieldMicrostructure}(g) -- and is accompanied by stress softening. The intensity of plastic activity increases with bands forming across grains till an applied strain of $\gamma = 0.04$, at which point twinning begins to nucleate -- Figure~\ref{fig:PhaseFieldMicrostructure}(d). As the loading continues increasing, existing twins grow, new twins appear, and the slip intensity grows. This twinning propagation is accompanied by further softening. Note that there are multiple minor load drops as micro-twins nucleate, but these quickly recover as the twins are pinned by either grain boundaries or plastic zones. Importantly both slip and twinning proceed as bands within favorable grains. Further, the twin boundaries are oriented according to the expected twin and reciprocal twin boundary orientations (indicated by gray and white arrows respectively in Figure~\ref{fig:PhaseFieldMicrostructure}(f)).

\paragraph{Propensity for slip and twinning}
The propensity, $p$, of the inelastic deformations is computed using the applied average deformation, $\bar{\mathbf{F}}$, local twin shear and normal directions, $\hat{\mathbf{b}}^{\, \text{t}}$ and $\hat{\mathbf{n}}^{\, \text{t}}$, and local slip shear and normal directions, $\hat{\mathbf{b}}^{\, \text{p}}$ and $\hat{\mathbf{n}}^{\, \text{p}}$, as follows:

\begin{equation}
    p = \Big[ \bar{\mathbf{F}} : \operatorname{sym}(\hat{\mathbf{b}}^{\, \text{t}} \otimes \hat{\mathbf{n}}^{\, \text{t}}) \Big]_+ - \left| \bar{\mathbf{F}} : \operatorname{sym}(\hat{\mathbf{b}}^{\, \text{p}} \otimes \hat{\mathbf{n}}^{\, \text{p}}) \right|.
\end{equation}

\noindent The operation $[\cdot]_+$ projects to the positive real axis, required to capture the asymmetry of twinning. The absolute value for the slip term captures its bidirectional behavior. The resulting values of $p \in (-1, 1)$ are plotted in Figure~\ref{fig:PhaseFieldMicrostructure}(a), with $-1$ indicating alignment with slip and $1$ corresponding to alignment with twinning. Comparing the propensity values in Figure~\ref{fig:PhaseFieldMicrostructure}(a) to the final twin and slip configuration, Figure~\ref{fig:PhaseFieldMicrostructure}(f) and (j), we observe both slip and twinning dominating in the respective dark and light grains. However, there are still instances where the less preferable system is present inside a grain, indicating the highly heterogeneous interaction between the inelastic and anisotropic elasticity.

\paragraph{Nucleation}

\begin{figure}
	\centering
    \includegraphics[width=\linewidth]{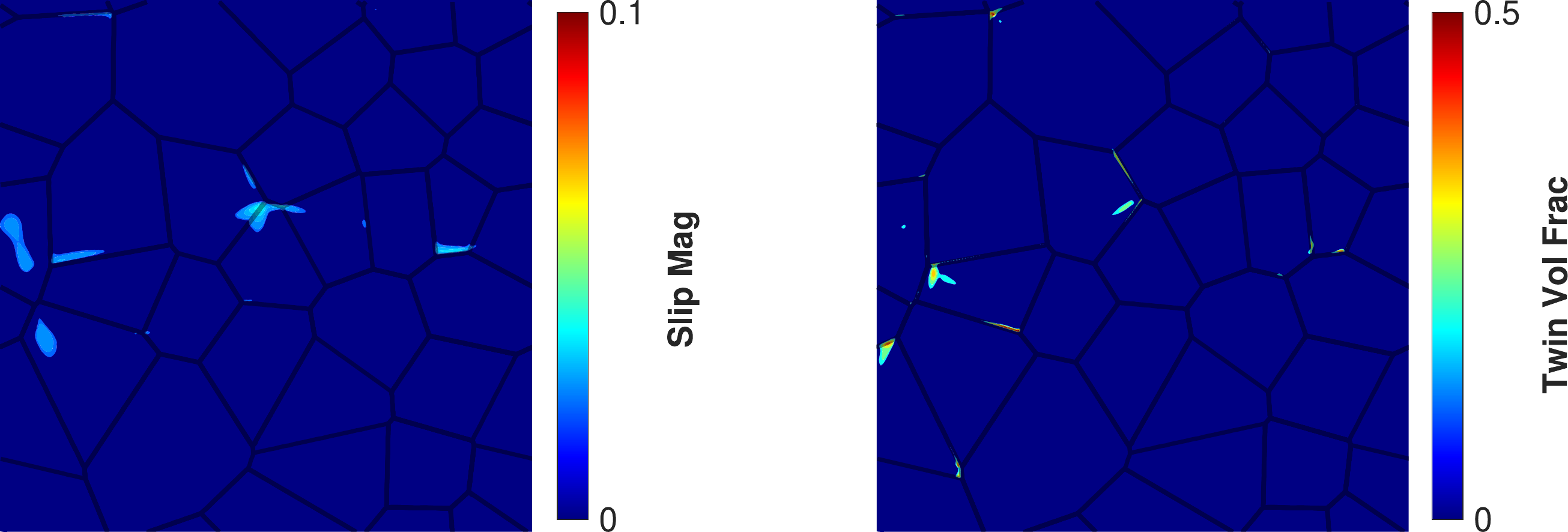}\\
	\caption{Nucleation. (a) Slip nucleation at $\gamma = 0.01$ (b) Twin nucleation at $\gamma = 0.04$.}
	\label{fig:TwinSlipNucleation}
\end{figure}

The anisotropy of elastic moduli in each grain gives rise to a heterogeneous distribution of elastic moduli across the polycrystalline specimen. Consequently, the state of stress in the initial elastic phase is highly heterogeneous with stress concentrations at grain boundaries. These stress concentrations lead to small amounts of slip occurring early in the loading. We see the first emergence of macroscopically significant slip at $\gamma = 0.01$; this is highlighted in Figure~\ref{fig:TwinSlipNucleation}(a) which reproduces the results of Figure~\ref{fig:PhaseFieldMicrostructure}(g) with a magnified scale. We notice that higher levels of slip are concentrated at triple junctions of grains with significant amounts of misalignment. In other words, slip nucleates near the triple junctions. This is also true for twinning, as seen in Figure~\ref{fig:TwinSlipNucleation}(b), which shows results with a magnified scale at a strain of $\gamma = 0.04$, a time shortly before Figure~\ref{fig:PhaseFieldMicrostructure}(d). Thus, the elastic anisotropy leads to stress-risers that enable nucleation at triple junctions.


\paragraph{Bridging}
Twinning can provide a bridge for a slip band to extend across an unfavorable grain. This is highlighted by a circle in Figure~\ref{fig:PhaseFieldMicrostructure}(a), (f), and (j). This provides further evidence of the nonlocal nature of morphology and the interplay between inelastic deformation mechanisms.

\begin{figure}
	\centering
    \includegraphics[width=\linewidth]{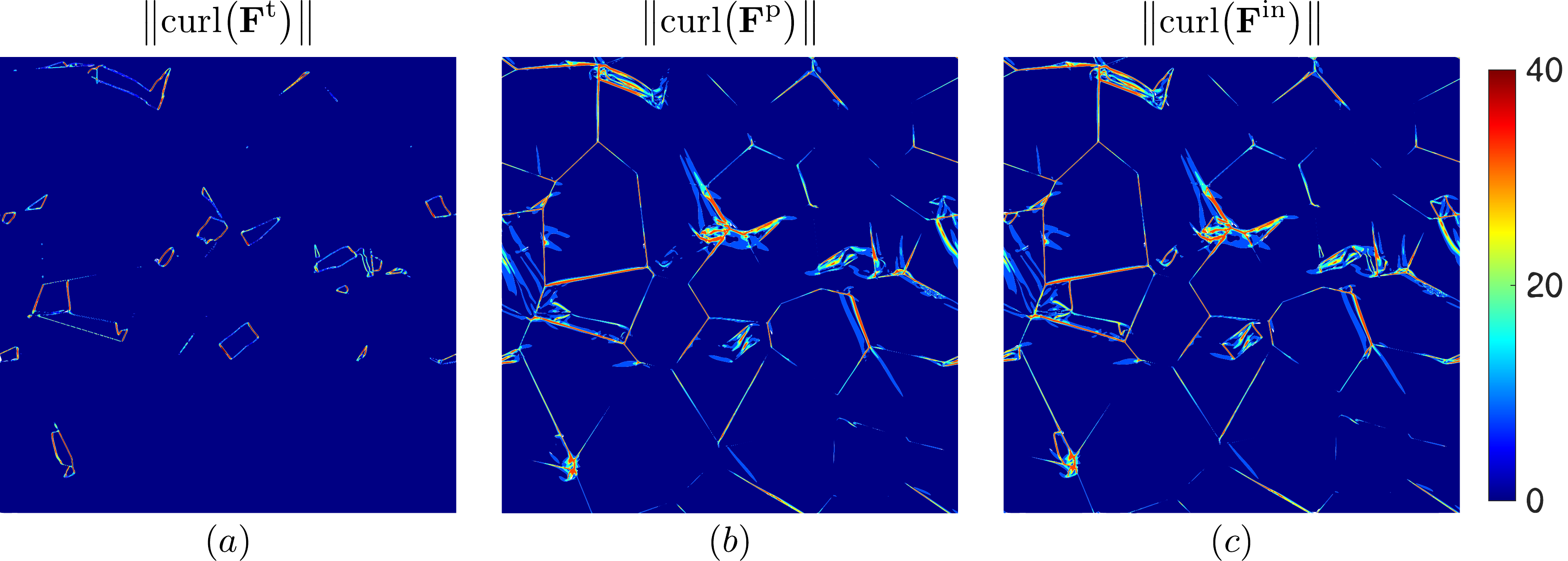}
	\caption{Geometrically necessary dislocation density. Magnitude of the curl of (a) twinning, (b) plastic slip, and (c) inelastic (plastic and twinning) deformations.}
	\label{fig:CurlF}
\end{figure}

\paragraph{Geometrically necessary dislocations}
The bridging across grains points to the the role of kinematic compatibility of the inelastic deformation in driving the cooperative interaction between grains. In plasticity, Nye's dislocation tensor or the curl of the plastic deformation gradient describes the geometrically necessary dislocations -- dislocations that are necessary to overcome the incompatibility of the plastic deformation \cite{Nye1953, Kaiser2019}. Figure~\ref{fig:CurlF} shows the magnitude of the curl of the slip, twinning, and combined inelastic deformations -- we may regard them as geometrically necessary slip, twinning, and inelastic dislocations. We see a large density at grain boundaries where twin/slip bands kink or where the twin bands are bridged by dislocations. We also see some twinning dislocations along twin boundaries as they may not be perfectly aligned.

\paragraph{Fluctuations}
\begin{figure}
	\centering
    \includegraphics[width=\linewidth]{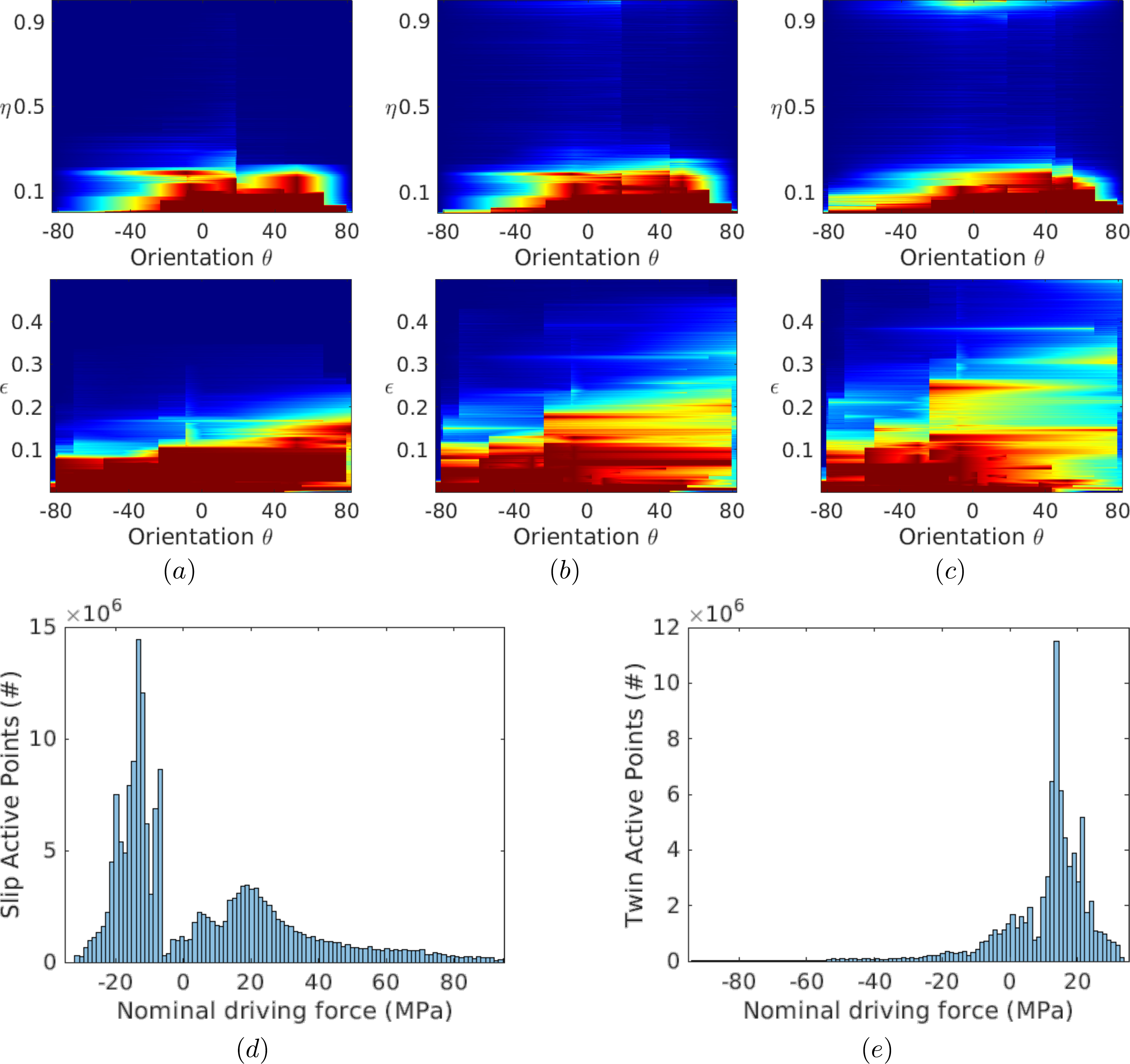}
	\caption{Statistical features of inelastic deformations for the Figure~\ref{fig:PhaseFieldMicrostructure} case. (a--c) Heat map showing the distribution of twin volume fraction and plastic activity as a function of grain orientation at strains of $\gamma = 0.01, 0.044, 0.1$ respectively. Blue represents little activation at that orientation and inelastic value, while red represents a larger number of material points with that orientation and inelastic value. (d--e) Cumulative histogram of twin and slip activity v.s. nominal driving force.}
	\label{fig:Statistics}
\end{figure}

The role of heterogeneity or fluctuations in slip and twin evolution is illustrated in Figure~\ref{fig:Statistics}. Figure~\ref{fig:Statistics}(a--c) displays heat maps of twin and slip activity versus local orientation, with the colors indicating the number of points active at that orientation and level of activity. We see large twin volume fractions and slip activity centered around favorable orientations, $22.5^\circ$ for twinning and both $22.5^\circ$ and $-22.5^\circ$ for slip. However, the activation for both spread across to more unfavorable grains. Additionally, we see a range of twin fraction and slip activity present at each orientation, indicating that the twin and slip activity can differ drastically in similarly orientated grains. Figure~\ref{fig:Statistics}(d) and (e) show the cumulative histogram of the active inelastic systems versus the nominal driving force, computed by projecting the macroscopic load onto the local orientation. The histograms illustrate that the activation of slip and twinning is centered ``near" their critical values, $\pm 4$MPa for slip and $9$MPa for twinning, though the peaks are shifted due to the rate dependence, hardening, and stress fluctuations. Further, we see the activation of various driving forces, even those close to zero or negative. These results illustrate that average values of micro-mechanical fields are insufficient at describing the complex behavior shown here.

\paragraph{Comparison with experimental observations}

These results are in qualitative agreement with various experimental observations in hcp materials.

Recently, Orozco-Caballero {\it et al.} \cite{Orozco-Caballero2017} experimentally found, using high-resolution digital image correlation and electron back-scatter diffraction imaging of polycrystal magnesium alloys, that the activation of twinning and unfavorable slip systems help accommodate strain incompatibility at grain boundaries between drastically differing grains.

The observations of specimen-spanning deformation bands dominating morphology and slip-twin bridging support these experimental findings on 

\subsection{Asymmetry of response}
\begin{figure}
	\centering
    \includegraphics[width=\linewidth]{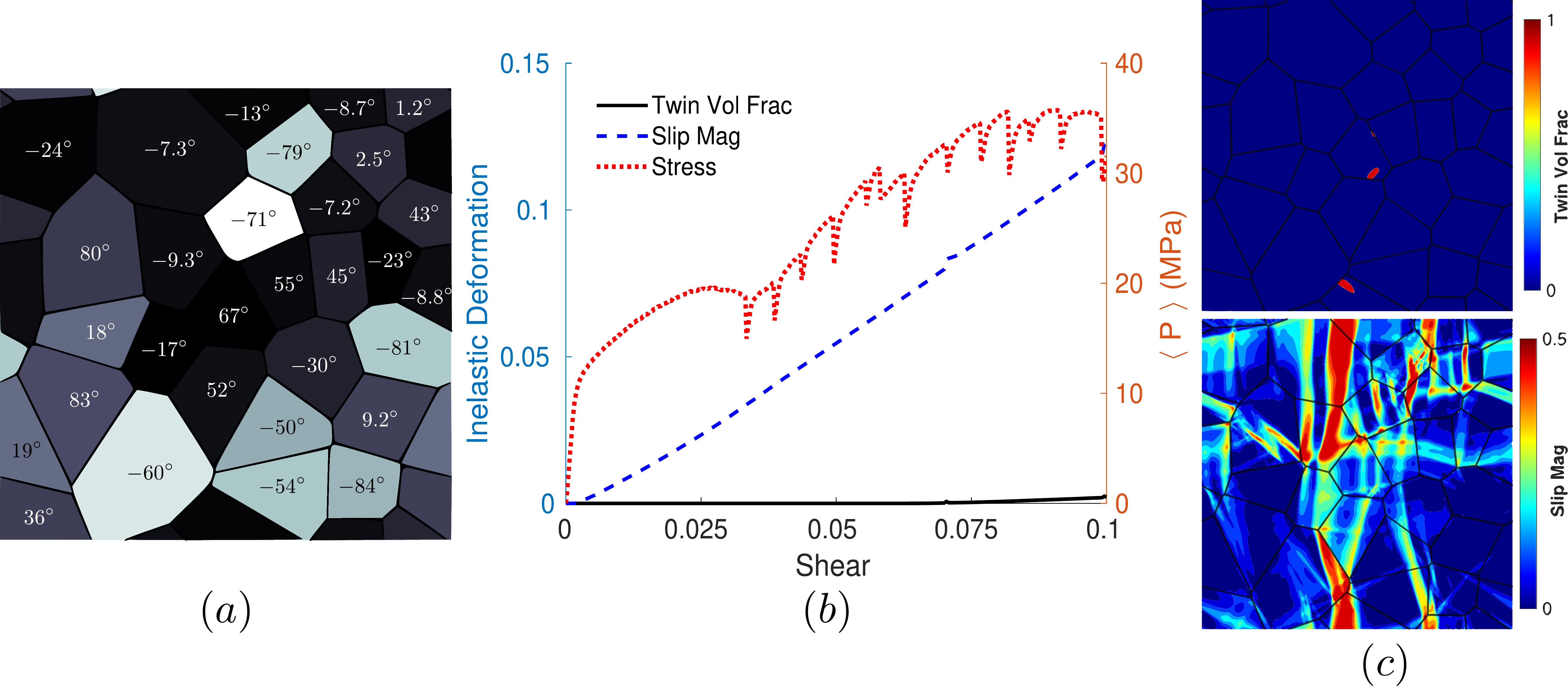}
	\caption{Asymmetry of response. Deformation of polycrystalline specimen subjected to reversed simple shear loading compared to Figure~\ref{fig:PhaseFieldMicrostructure}. (a) The polycrystalline specimen. The grayscale shows Schmidt factors for twinning and slip. Darker grains are favorably oriented for slip and lighter grains for twinning, while the numbers indicate the grain orientation. (b) Stress-strain behavior and evolution of average slip magnitude and twin volume fraction. (c) Snapshot of twin volume fraction ($\eta$) and slip magnitude ($\gamma^{\, \text{p}}$) at the final strain of $\gamma = -0.1$.}
	\label{fig:ReverseLoading}
\end{figure}

Starting with the same specimen as in Figure~\ref{fig:PhaseFieldMicrostructure}, we apply a shear in the opposite direction. The resulting twin and slip morphology, stress-strain curve, and average inelastic values are in Figure~\ref{fig:ReverseLoading}. Compared to the forward loading direction, we see much less twin activity and differing locations of twinning. Similarly, the slip activity changes drastically, with different grains exhibiting large plastic deformation. Finally, the changes in local inelastic deformations result in a macroscopic stress-strain curve with more stress hardening and stress drops. All of this is the analog in shear of the well-known tension-compression anisotropy in hcp materials.

\subsection{Proportional loading-unloading-reverse loading}

\begin{figure}
	\centering
    \includegraphics[width=\linewidth]{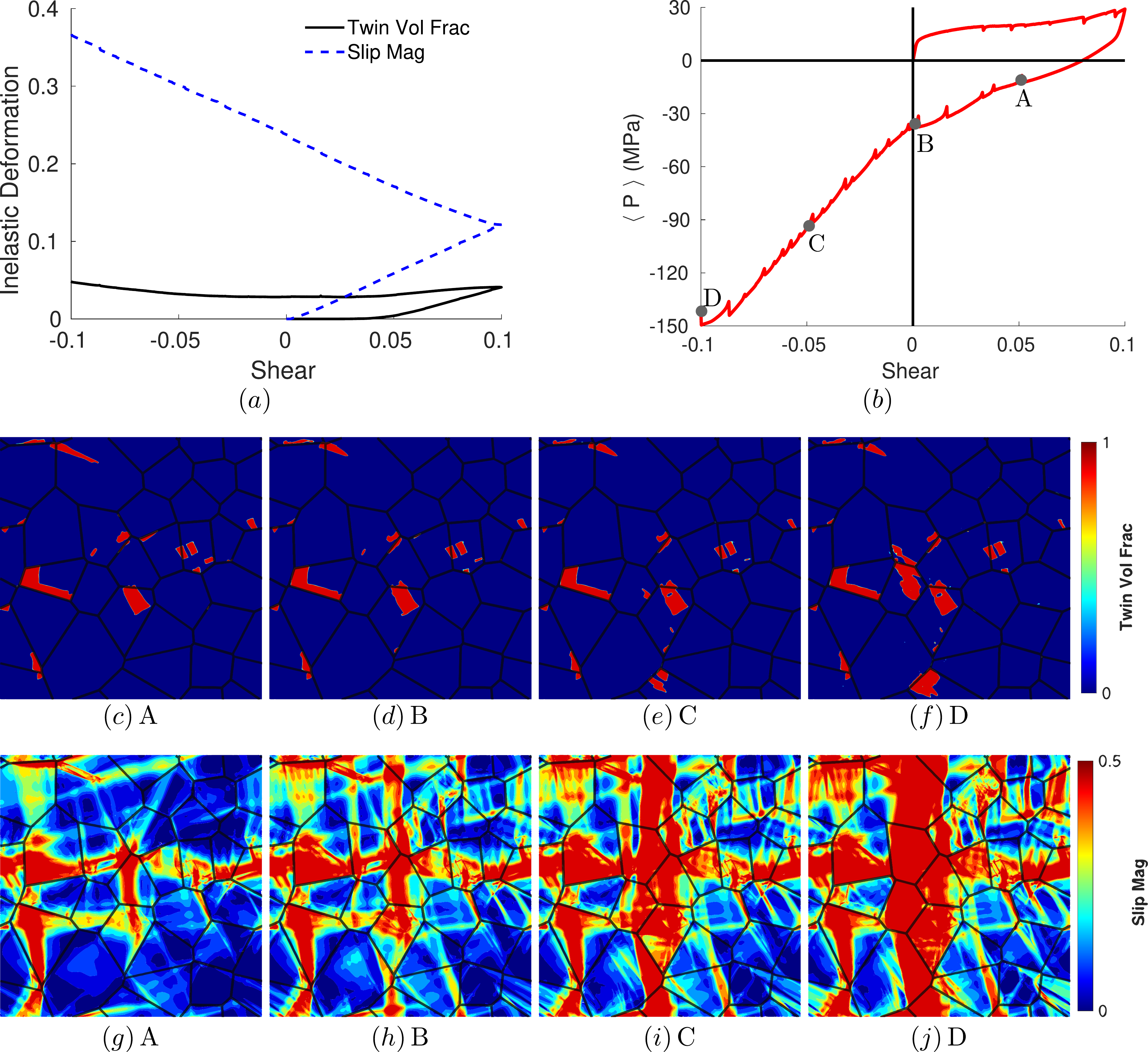}\\
	\caption{Proportional forward-reverse deformation. (a) Average twin volume fraction and slip magnitude versus shear strain. (b) Average stress-strain plot with markers indicating points at which microstructure plots are obtained. (c--f) Twin volume fractions at points during the reverse loading. (g--j) Slip magnitudes at points during the reverse loading.}
	\label{fig:MicroForwardReverse}
\end{figure}
Once again, starting with the same specimen as in Figure~\ref{fig:PhaseFieldMicrostructure}, we apply a shear in the forward direction, and then reverse the direction of shear till we have sheared it in the opposite direction. The results are shown in Figure~\ref{fig:MicroForwardReverse}. The forward shearing is as before. As the shear direction is reversed, we see significant residual strains. Further, there is a small amount of detwinning, though it is not complete, while the slip intensity continues increasing. Note that if we did not have slip, the detwinning would be significant. Thus, slip induced deformation pins twins in place. As the applied shear goes through zero and eventually increases in the reverse direction, the twin volume fraction morphology remains relatively constant till a large reverse shear, while the slip continues growing. We also see significant hardening. In particular, comparing Figures~\ref{fig:ReverseLoading} and \ref{fig:MicroForwardReverse} at an applied shear of $-0.1$, we see that the stress is significantly higher and twin and slip morphology are drastically different as a result of the prior deformation. Thus, the interplay between slip and twinning provides a significant complexity in the role of prior deformation.

Supplementary Figures S9 and S10 provide additional proportional loading examples.

\subsection{Non-Proportional Loading}
\begin{figure}
	\centering
    \includegraphics[width=\linewidth]{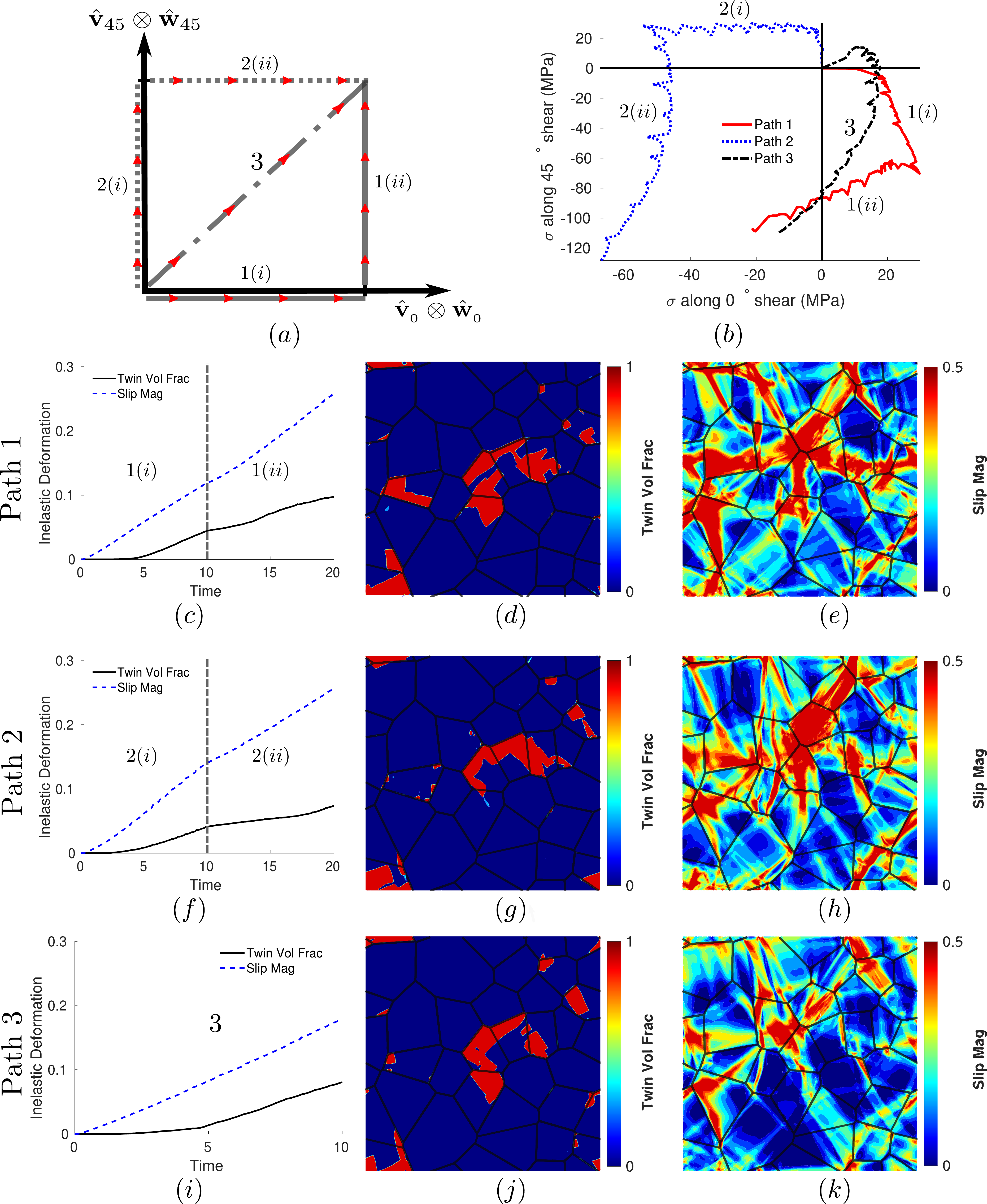}\\
	\caption{Non-proportional loading. Three macroscopic strain paths (a) are simulated with identical final strain. (b) Shear stress paths for each loading. Average twin volume fraction and slip magnitude and final twin and slip morphology for path 1 (c--e), path 2 (f--h), and path 3 (i--k).}
	\label{fig:NonProportional2_45}
\end{figure}

Non-proportional loading is a crucial loading case for the failure analysis of materials and more closely matches applications. We consider two strain directions -- shear $\mathbf{I} + \hat{\mathbf{u}}_0 \otimes \hat{\mathbf{w}}_0$ along $0^\circ$ and shear $\mathbf{I} + \hat{\mathbf{u}}_{45} \otimes \hat{\mathbf{w}}_{45}$ along $45^\circ$ where 
\begin{equation}
\hat{\mathbf{u}}_0 = \{1,0\}, \ \hat{\mathbf{w}}_0 = \{0,1\}; \quad 
\hat{\mathbf{u}}_{45} = 1/\sqrt{2} \{1,1\}, \ \hat{\mathbf{w}}_{45} = 1/\sqrt{2} \{1,-1\}.
\end{equation} 
We consider three strain paths -- first shearing along $0^\circ$ and then along $45^\circ$, first shearing along $45^\circ$ and then along $0^\circ$, and a combined shear path -- where the end macroscopic strain states are the same. The results are shown in Figure~\ref{fig:NonProportional2_45}. We see that the end state of stress is different, as are the twin and slip morphologies. Interestingly, load paths 1 and 3 lead to a similar (though distinct) state of stresses though the slip and twin morphologies are different. Path 2 leads to significantly higher hardening. This example again shows the complex history dependence of the state of stress and morphology.

Supplementary Figures S11 and S12 provide additional non-proportional loading examples.

\subsection{Twin and slip activity}
We explore the variation of twin and slip activity for differing nucleation threshold values by varying the parameters, $\tau^{\, \text{p}}_0$, $\tau^{\, \text{t}}_0$, and $M$, all of which impact the nucleation and propagation of twinning and slip. To isolate the effect of these material parameters, we fix the grain structure from Figure~\ref{fig:PhaseFieldMicrostructure}a. The resulting average stress, slip magnitude, and twin volume fraction are shown below in Figure~\ref{fig:ParameterStudy}; the final twin and slip morphologies are provided in supplementary Figures S13--S15.

\begin{figure}
	\centering
    \includegraphics[width=\linewidth]{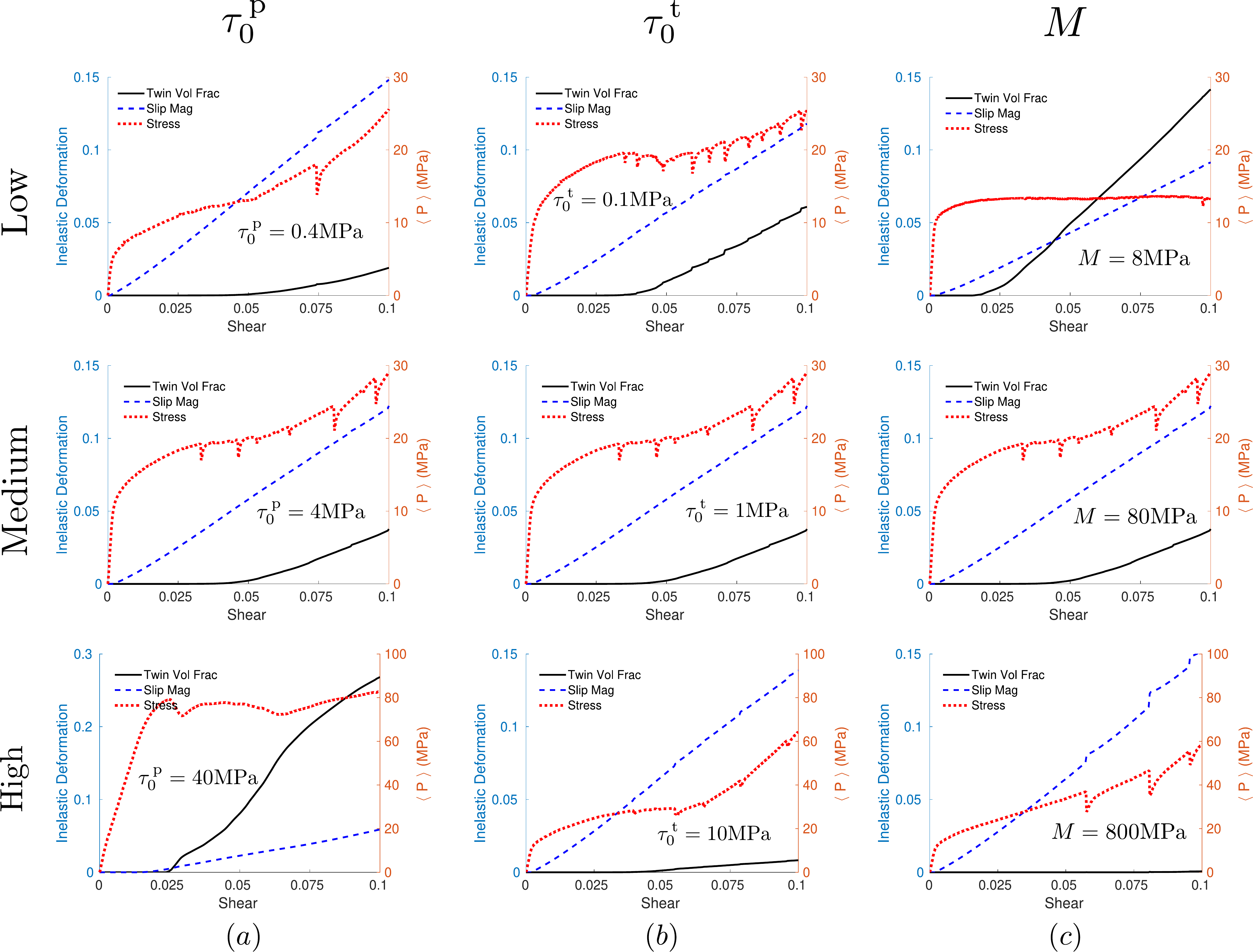}
	\caption{Role of parameters on twin and slip activity. Average stress, twin volume fraction and slip magnitude for varying $\tau^{\, \text{p}}_0$ in (a), $\tau^{\, \text{t}}_0$ in (b) and $M$ in (c). All other parameters as before.}
	\label{fig:ParameterStudy}
\end{figure}

\paragraph{Surface energy}
As mentioned in Section~\ref{sec:TwinNucleation} the double-well parameter, $M$, controls twin nucleation and is supported by earlier non-zero twin volume fractions for smaller $M$ in Figure~\ref{fig:ParameterStudy}. This earlier twin nucleation leads to larger final twin volume fractions and a smaller final slip magnitude for small $M$. Lastly, the stress-strain curves exhibit an earlier twin yield and more stress softening for small $M$ values. For larger values of $M$, the opposite is found, less (essentially negligible) twin volume fractions, more slip activity, and more stress hardening, as seen in Figure~\ref{fig:ParameterStudy}.


\paragraph{Twin rate hardening}
The rate hardening parameter, $\tau_0^{\, \text{t}}$, affects both twin nucleation and propagation. As $\tau_0^{\, \text{t}}$ decreases, the twin nucleation change is relatively minor, but there is a more noticeable increase in the final twin volume fraction. Consequently, with the decreasing $\tau_0^{\, \text{t}}$, the slip magnitude decreases due to the higher presence of twinning. Lastly, the stress-strain curve exhibits a slightly earlier twin yield and more stress softening for lower values of $\tau_0^{\, \text{t}}$. For larger values of $\tau_0^{\, \text{t}}$, the opposite is found, more slip activity, smaller average twin volume fraction, and more stress hardening.


\paragraph{Slip rate hardening}
The rate hardening parameter, $\tau_0^{\, \text{p}}$, affects both slip nucleation and propagation. For smaller $\tau_0^{\, \text{p}}$ values, the average slip magnitude shows earlier slip nucleation and higher final slip activity, while the average twin volume fraction shows decreased twin presence. Lastly, the stress-strain curve exhibits an earlier slip yield point and larger hardening due to the suppression of twins for smaller $\tau_0^{\, \text{p}}$. For larger values of $\tau_0^{\, \text{p}}$ the opposite is found, less slip activity, more twinning activity, and a delayed slip yield point.

%

\subsection{Texture}
We consider the same grain structure, material parameters, and loading as Section \ref{sec:typical} but with different grain orientations.

\begin{figure}
	\centering
    \includegraphics[width=\linewidth]{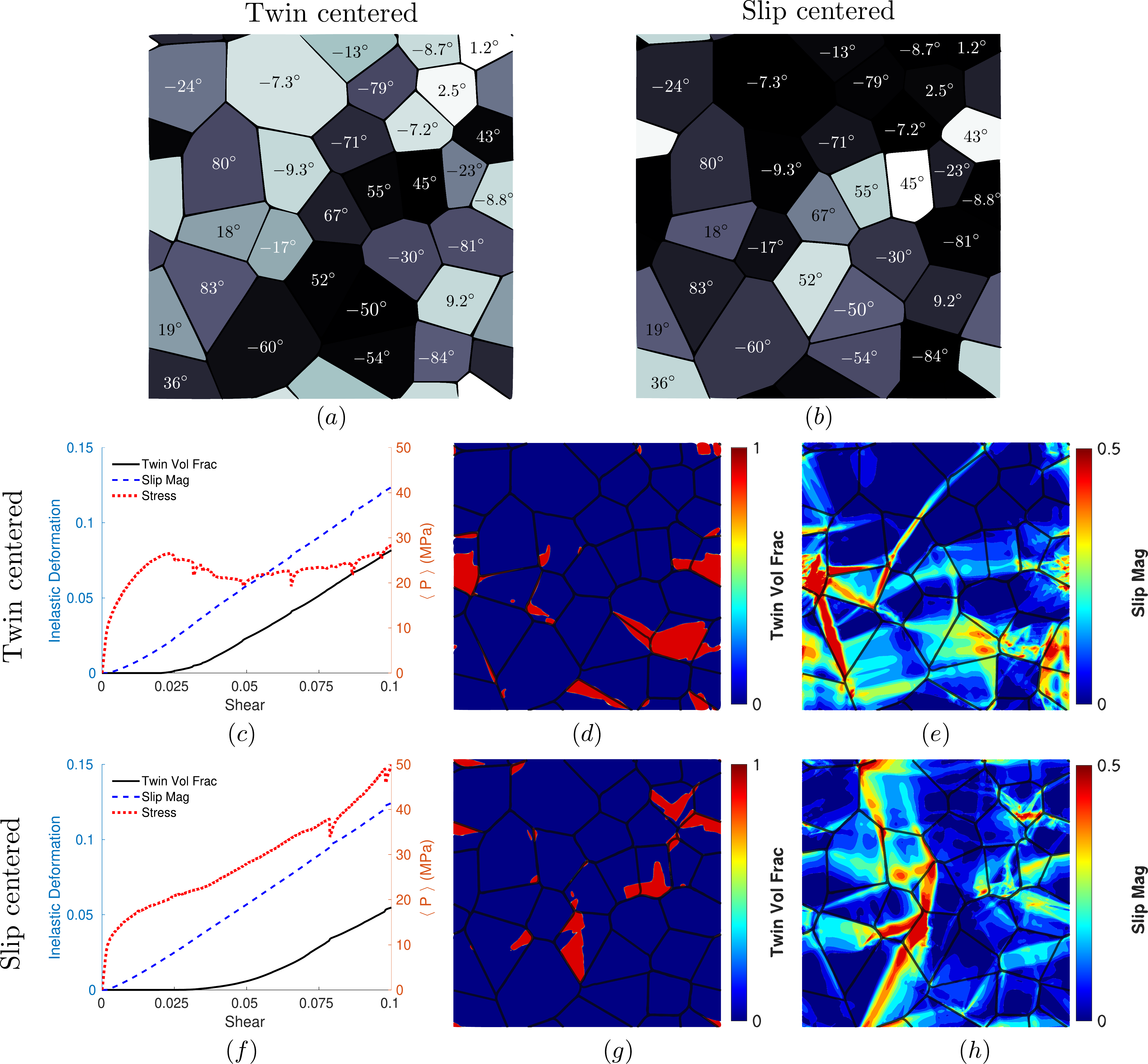}
	\caption{Role of Texture. Deformation of the different polycrystalline specimens subjected to simple shear. Polycrystalline specimen for twin centered (a) and slip centered (b) grain angles. The grayscale shows Schmidt factors for twinning and slip. Darker grains are favorably oriented for slip and lighter grains for twinning, while the numbers indicate the grain orientation. Stress-strain behavior and average inelastic deformations for twin centered (c) and slip centered (f) grains. Final twin volume fraction ($\eta$) and slip magnitude ($\gamma^{\, \text{p}}$) at a final strain of $\gamma = 0.1$ for twin centered (d--e) and slip centered (g--h) grains.}
	\label{fig:TextureRole}
\end{figure}

\paragraph{Twin centered texture}
The grain angles are now sampled from a Gaussian centered around the twin direction -- seen in Figure~\ref{fig:TextureRole}(a). The resulting final twin and slip morphology, stress-strain curves, and average inelastic deformations are shown in Figure~\ref{fig:TextureRole}(c--e). Compared to the original case, we see more twin activity and its presence in different grains due to the change in favorable grains. Similarly, the slip activity changes drastically with differing grains exhibiting large plastic deformation. Finally, the increased presence of twin bands leads to more stress drops in the stress-strain curve.

\paragraph{Slip centered texture}
The grain angles are now sampled from a Gaussian about the slip direction -- seen in Figure~\ref{fig:TextureRole}(b). The resulting final twin and slip morphology, stress-strain curve, and average inelastic deformations are shown in Figure~\ref{fig:TextureRole}(e--h). Compared to the original case, we see less twin activity and its presence in different grains due to the change in favorable grains. Similarly, the slip activity changes drastically with more grains exhibiting large plastic deformation. Finally, the decrease in twin bands leads to more hardening in the stress-strain curve.

\subsection{Comparison with the results of a pseudo-slip model}
We now compare the results above with those obtained from a pseudo-slip model following \cite{Chang2015}.  The details are in the appendix and the implementation is similar to that used for our model.

\begin{figure}
	\centering
    \includegraphics[width=0.86\linewidth]{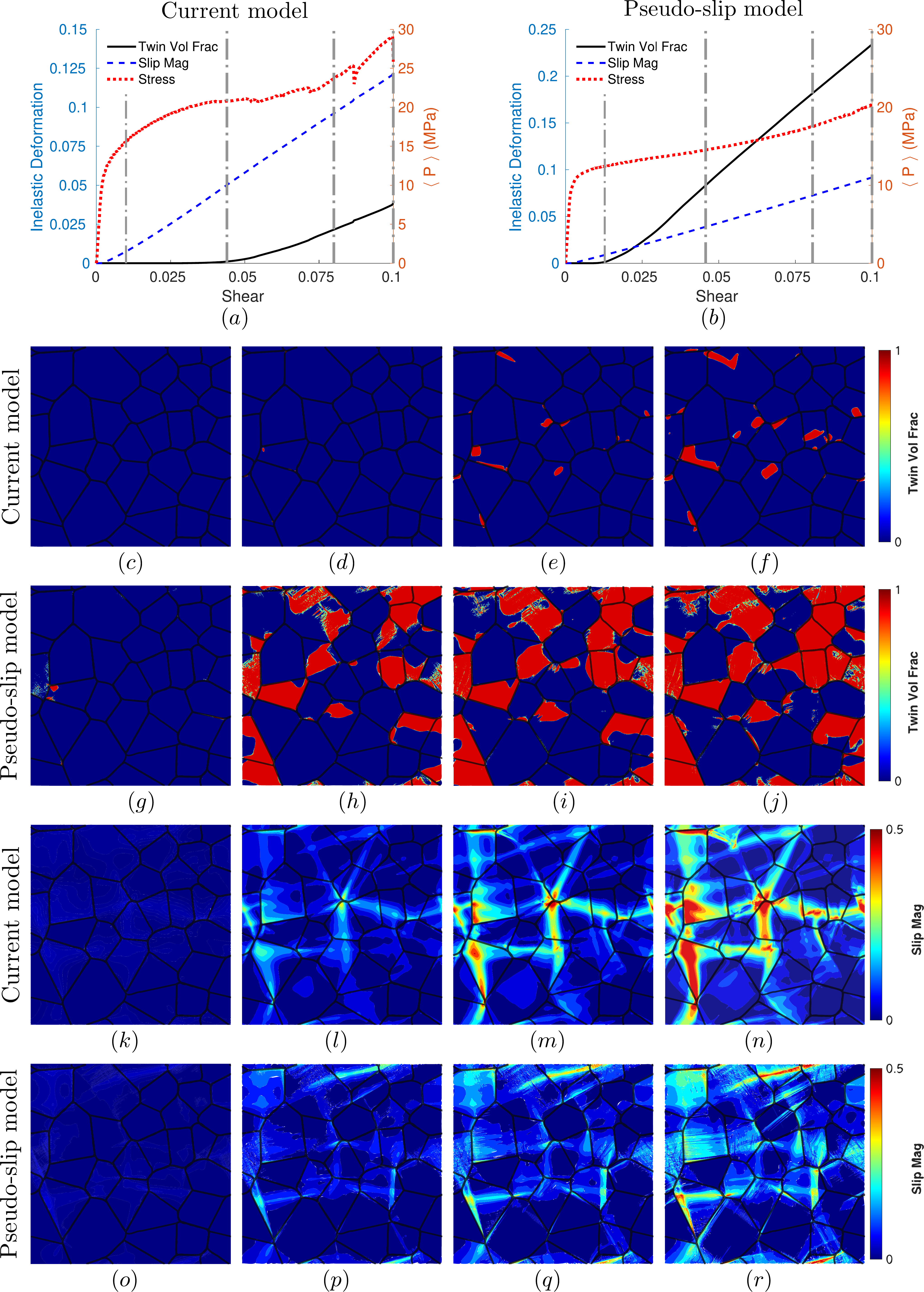}\\
	\caption{Comparison with pseudo-slip model. Stress strain curve and average slip magnitude and twin volume fraction for phase-field (a) and pseudo slip (b). The gray dashed lines correspond to points where the microstructure is plotted. Phase-field's twin volume fraction (c--f) and pseudo slip's twin volume fraction (g--j) at strains of $\gamma = 0.01, \, 0.044, \, 0.07, \, 0.1$. Phase-field's slip magnitude (k--n) and pseudo slip's twin volume fraction (o--r) at strains of $\gamma = 0.01, \, 0.044, \, 0.07, \, 0.1$.}
	\label{fig:PseudoSlipComp}
\end{figure}

Figure~\ref{fig:PseudoSlipComp} compares the results of the two models under identical parameters, initial microstructure, and loading. Figure~\ref{fig:PseudoSlipComp}(b) shows the pseudo-slip stress-strain curve has only one yield point due to slip and twinning now having identical nucleation criteria. Comparing the evolution of the average volume fractions and slip activity (Figure~\ref{fig:PseudoSlipComp}(a) and (b)), we see that twinning nucleates early and dominates the deformation in the pseudo-slip model compared to our model. Indeed, comparing Figure~\ref{fig:PseudoSlipComp}(c--f) to Figure~\ref{fig:PseudoSlipComp}(g--j), we see early and extensive twinning in the pseudo-slip model with drastically different morphology. Our model predicts well-defined twin bands, while the pseudo-slip model predicts diffuse twinning extending across the grains. All of this is expected since the pseudo-slip model does not seek to define thin twins, does not account for twin boundary energy that sets a length scale, and does not have any nucleation barrier. Finally, Figure~\ref{fig:PseudoSlipComp}(k--m) and Figure~\ref{fig:PseudoSlipComp}(n--q) show that slip is greatly suppressed in the pseudo-slip model. This slip reduction is due to twinning accommodating a greater portion of the deformation and diffuse twins contributing smaller incompatibilities.

\section{Conclusion and Discussion}\label{sec:Conclusion}

In this work, we have presented a model that describes deformation twins and plastic dislocation slip at the scale of multiple grains, their morphology, their interaction, and their implications on macroscopic behavior. We start with the detailed kinematics of twinning following Mahajan and Christian \cite{Christian1995}, and implement it using a phase-field framework to incorporate detailed morphology, nucleation barrier, surface energy, propagation drag, and rate hardening. We treat dislocation plasticity in the framework of crystal plasticity. We implement the model on graphical processing units following \cite{Zhou2020}. The goal of this paper is to understand various aspects of the interaction between deformation twinning and dislocation slip in polycrystal domains. Therefore we use detailed simulations in two dimensions to study nucleation and growth of twins, the evolution of slip, the cooperative mechanism of bridging across grains, geometrically necessary dislocations and the role of fluctuations.

We observe that macroscopic quantities like propensity for twinning and slip, and nominal driving force (Schmid factor) are suggestive but not predictive of the presence of twinning and slip in a grain.  Grains well-oriented for slip were more likely to slip and those oriented for twinning were more likely to twin (see Figure \ref{fig:PhaseFieldMicrostructure}).   However, grains with the similar orientation behave may behave differently, and one can observe twinning and slip even when the nominal driving force or Schmidt factor is negative (Figure \ref{fig:Statistics}).   This is in agreement with observations in magnesium and its alloys \cite{GodetMgTwins,beyerlein_2010,jonas2011role}, and zirconium and its alloys \cite{BingertZrTwins02,CapolungoTwinning2009,abdolvand2013,lind_2014}.  

There are two notable reasons for this behavior.  The first is the anisotropy of the elastic modulus that leads to a heterogeneous state of stress.  In particular, we have significant amounts of stress concentrations at triple junctions, which in turn leads to heterogeneous nucleation (see Figure \ref{fig:TwinSlipNucleation}).  The second is incompatibility between neighboring grains.  This is evident both during nucleation where it tends to occur in multiple grains, as well as during growth where deformation mechanisms can bridge poorly oriented grains to connect well oriented grains (see the circled regions in Figure \ref{fig:PhaseFieldMicrostructure}).  This is highlighted by examining the geometrically necessary dislocations (Figure \ref{fig:CurlF}).  These are again in qualitative agreement with experimental observations.  In situ electron back scatter diffraction (EBSD) observations of Guo {\it et al.} \cite{guo_2014} in magnesium emphasize the simultaneous nucleation of twins in multiple grains and and the formation of twin chains.  Combined high-resolution digital image correlation (HRDIC) and EBSD observations of Orozco-Caballero {\it et al.} \cite{Orozco-Caballero2017} in the magnesium alloy AZ31 show the heterogeneous nature of deformation at the sub-granular and multi-granular scales, and how hard slip can be activated to bridge deformation across grains.  Similarly, EBSD observations of AZ31 of Jonas {\it et al.} \cite{jonas2011role} confirmed that twins formed in low Schmidt factor grains to bridge those formed in high Schmidt factor grains.  

We also used simulations to study the asymmetric deformation response due to twinning, and the history dependence in both proportional and non-proportional loading.   All of these provide insights into the deformation behavior of low symmetry crystals where both deformation twinning and dislocation slip are significant.

We then compare our model with previous work that treated twinning as `pseudo-slip' in the framework of crystal plasticity.   A key observation is that the pseudo-slip significantly over-predicts the extent of twinning.  This is consistent with the detailed comparative study between (EBSD) observations and pseudo slip based crystal plasticity simulations by Abdolvand and Daymond \cite{abdolvand2013}; this study showed that the pseudo-slip model significantly over-predicted the extent of twinning.  The pseudo-slip approach ignores twin-boundary energy and thus lacks a length-scale leading to an under-prediction of the nucleation barrier.

The establishment of the current model and the detailed phenomenological studies enables further development. The next step is to implement this model in three dimensions with multiple slip and twin systems and to conduct a detailed comparison with experimental observations in hcp metals like magnesium. An important development in this direction is to include the transmutation of slip at twin boundaries based on the results of atomistic simulations.

\section*{Acknowledgement}
We gratefully acknowledge the support of the US Army Research Laboratory through Cooperative Agreement Number W911NF-12-2-0022 and the US National Science Foundation through ``Collaborative Research: Optimal Design of Responsive Materials and Structures'' (DMS-2009289). The views and conclusions contained in this document are those of the authors and should not be interpreted as representing the official policies, either expressed or implied, of the Army Research Laboratory or the U.S. Government. The U.S. Government is authorized to reproduce and distribute reprints for Government purposes notwithstanding any copyright notation herein. The computations presented here were conducted on the Resnick High Performance Cluster at the California Institute of Technology.


\renewcommand{\thesection}{\mbox{Appendix-\Alph{section}}}
\setcounter{section}{0}

\section{Numerical method}

We discretize the equilibrium equation (\ref{eq:equil}) and evolution equations (\ref{eq:SlipStationary} and \ref{eq:TwinStationary}) in time using an implicit time discretization, and rewrite them as an incremental variational principle (e.g. \cite{Ortiz1999}). The increment in deformation, twin and slip are obtained as the solution to the variational problem, 
\begin{equation}
{\mathbf{y}}_{\, \text{n+1}}, \eta_{\, \text{n+1}},  \gamma^{\, \text{p}}_{\, \text{n+1}} = \text{arg min} \int_\Omega \bigg( W(\nabla \mathbf{y}, \eta, \nabla \eta, \epsilon^{\, \text{p}},{\mathbf X}) + \Delta t\Psi^*\bigg( \frac{\Delta\gamma^{\, \text{p}}}{\Delta t}, \frac{\Delta\eta}{\Delta t} \bigg)  \bigg) d\Omega, \label{eq:InfImplicit}
\end{equation}
and the increment in plastic strain is 
\begin{equation}
\mathbf{F}^{\, \text{p}}_{\, \text{n+1}} =  \big(\mathbf{I} + \Delta\gamma^{\, \text{p}}\hat{\mathbf{b}}^{\, \text{p}}\otimes\hat{\mathbf{n}}^{\, \text{p}} \big)\mathbf{F}^{\, \text{p}}_{\, \text{n}}.
\end{equation}

We solve the variational problem following the accelerated computation micromechanics approach \cite{Zhou2020}. The basic idea is to use both deformation $\mathbf{y}$  and its gradient $\mathbf{F}$ as independent variables, and to similarly use both the twin phase-field, relabeled $\beta$, and its gradient $\nabla\eta$ as independent variables. We then treat the compatibility between the deformation and twin phase-field and their gradients as a constraint that is implemented using an augmented Lagrangian. So we consider the functional
\begin{equation}
\int_\Omega \bigg( W(\mathbf{F}, \beta, \nabla \eta, \epsilon^{\, \text{p}},{\mathbf X}) + W_{\, \lambda}(\mathbf{F}, \nabla\mathbf{y}, \beta, \eta) + \Delta t\Psi^*\bigg( \frac{\Delta\gamma^{\, \text{p}}}{\Delta t}, \frac{\Delta\beta}{\Delta t} \bigg)  \bigg) d\Omega, \label{eq:StatAL}
\end{equation}
where 
\begin{equation}
	W_{\, \lambda}(\mathbf{F}, \mathbf{y}, \eta, \beta) = \mathbf{\lambda}_F:(\mathbf{F} - \nabla\mathbf{y}) + \mathbf{\lambda}_\eta(\beta - \eta) + \frac{\rho_F}{2}\left\lVert \mathbf{F} - \nabla\mathbf{y}\right\rVert^2 + \frac{\rho_\eta}{2}\left\lVert \beta - \eta\right\rVert^2. \label{eq:ALEnergy}
\end{equation}
Here, $\rho_F>0, \rho_\beta>0$ are fixed constants. We have a saddle point problem where we minimize the functional over $ {\mathbf F}, {\mathbf y}, \eta, \beta$ and maximize over the Lagrange multipliers $\lambda_F, \lambda_\eta$.   We do so using the alternating direction method of minimizers:

At the (n+1)th time step, given the previous solution, $\mathbf{F}^n$, $\mathbf{y}^n$, $\eta^n$, $\beta^n$, $\gamma^n$, $\mathbf{\lambda}_F^n$, and $\mathbf{\lambda}_\eta^n$, and the current macroscopic strain $\mathbf{\bar{F}}^{n+1}$ the new equilibrium solution is given by iterating over $i$ in the following nested loops until convergence,

\begin{itemize}
    \item \textit{Step 1: ADMM for twinning.} Solve for $\beta_{i+1}$, $\eta_{i+1}$, and $\lambda_{\eta, i+1}$ while fixing $\mathbf{F}_i$ and $\gamma^p_i$
    
    \begin{itemize}
        \item \textit{Step 1a: Local Problem.} Update $\beta_{j+1}$ at each $\mathbf{X}$ by solving
            \begin{equation}
                W_\beta(\mathbf{F}_i,\beta_{j+1},\nabla\eta_{j},\epsilon^{\, \text{p}}_i,\mathbf{X}) + \Delta t\Psi^*_\beta\Big(\frac{\gamma^{\, \text{p}}_i - \gamma^{\, \text{p}, \, n}}{\Delta t}, \frac{\beta_{j+1} - \beta^{\, n}}{\Delta t}\Big) + \lambda_{\eta, \, j} + \rho_\eta(\beta_{j+1} - \eta_j) = 0
            \end{equation}
            
        \item \textit{Step 1b: Helmholtz projection.} Update $\eta_{j+1}$ by solving the partial differential equation
            \begin{equation}
                \nabla^2\eta_{j+1} - \frac{\rho_\eta}{\alpha}\eta_{j+1} = -\frac{1}{\alpha}\Big(\lambda_{\eta, \, j} + \rho_\eta\beta_{j+1} \Big)
            \end{equation}
        \item \textit{Step 1c: Update Lagrange multiplier.} Update $\lambda_{\eta, \, j+1}$
        \begin{equation}
            \lambda_{\eta, \, j+1} = \lambda_{\eta, \, j} + \rho_\eta(\beta_{j+1} - \eta_{j+1})
        \end{equation}
        \item \textit{Step 1d: Check for convergence.} Define the primal and dual feasibility as
        \begin{equation}
            r_{\eta, \, p} = \left\lVert \beta_{j+1} - \eta_{j+1}\right\rVert_{L^2}, \qquad r_{\eta, \, d} = \rho_\eta/M \left\lVert \eta_{j+1} - \eta_{j}\right\rVert_{L^2}.
        \end{equation}
        If $r_{\eta, \, p} \leq r^{\, \text{tol}}_p$ and $r_{\eta, \, d} \leq r^{\, \text{tol}}_d$, then $\beta_{i+1} = \beta_{j+1}$, $\eta_{i+1} = \eta_{j+1}$, and $\lambda_{\eta, i+1} = \lambda_{\eta, j+1}$ else return to \textit{Step 1a}.
    \end{itemize}
    
    \item \textit{Step 2: Local Problem for $\mathbf{F}$.} Update $\mathbf{F}_{i+1}$ and $\gamma^{\, \text{p}}_{i+1}$ at each $\mathbf{X}$ by solving
    \begin{align}
        &W_F(\mathbf{F}_{i+1},\beta_{i+1},\nabla\eta_{i},\epsilon^{\, \text{p}}_{i+1},\mathbf{X}) + \lambda_{F, \, i} + \rho_F(\mathbf{F}_{i+1} - \nabla\mathbf{y}_{i}) = 0\\
        \vspace{.5cm}
        &W_{\gamma^{\, \text{p}}}(\mathbf{F}_{i+1},\beta_{i+1},\nabla\eta_{i},\epsilon^{\, \text{p}}_{i+1},\mathbf{X}) + \Delta t\Psi^*_{\gamma^{\, \text{p}}}\Big(\frac{\gamma^{\, \text{p}}_{i+1} - \gamma^{\, \text{p}, \, n}}{\Delta t}, \frac{\beta_{i+1} - \beta^{\, n}}{\Delta t}\Big) = 0
    \end{align}
    \item \textit{Step 3: Helmholtz projection.} Update $\mathbf{y}_{i+1}$ by solving the partial differential equation
    \begin{equation}
        \nabla^2\mathbf{y}_{i+1} = \nabla\cdot \Big( \mathbf{F}_{i+1} + \frac{\lambda_{F,\, i}}{\rho_F} \Big)
    \end{equation}
    \item \textit{Step 4: Update Lagrange multiplier.} Update $\lambda_{F, \, i+1}$,
        \begin{equation}
            \lambda_{F, \, i+1} = \lambda_{F, \, i} + \rho_F(\mathbf{F}_{i+1} - \nabla\mathbf{y}_{i+1}).
        \end{equation}
    \item \textit{Step 5: Check for convergence.} Define primal and dual feasibility
    \begin{equation}
        r_{F, \, p} = \left\lVert \mathbf{F}_{i+1} - \nabla\mathbf{y}_{i+1}\right\rVert_{L^2}, \qquad r_{F, \, d} = \rho_F/\mu \left\lVert \nabla\mathbf{y}_{i+1} - \nabla\mathbf{y}_{i}\right\rVert_{L^2}.
    \end{equation}
    If $r_{f, \, p} \leq r^{\text{tol}}_p$ and $r_{f, \, d} \leq r^{\text{tol}}_d$, then $\mathbf{F}^n = \mathbf{F}_{i+1}$, $\mathbf{y}^n = \mathbf{y}_{i+1}$, $\beta^n = \beta_{i+1}$, $\eta^n = \eta_{i+1}$, $\lambda^n_F = \lambda_{F, i+1}$ and $\lambda^n_\eta = \lambda_{\eta, i+1}$ else return to \textit{Step 1}.
\end{itemize}
\vspace{\baselineskip}

The algorithm is known to converge for sufficiently large $\rho_F, \rho_\beta$, though too large values slows down the convergence. We use periodic boundary conditions and fast Fourier transforms to solve the Helmholtz project (steps 1b and step 3). This algorithm is known to show good performance relative to other FFT-based algorithms but has the major benefit of easy parallel implementation. Steps 1a and 2 are local nonlinear problems that are solved in parallel using steepest gradient descent or Newton-Raphson methods. Steps 1b and 3 are Helmholtz projections for which there are efficient parallel algorithms. And Steps 1c, 1d, 4, and 5 are simple updates and checks. Thus the entire iterative algorithm is implemented in parallel using Graphical Processing Units (GPUs), which provide thousands of cores for accelerated computations, allowing us to run large-scale simulations for more refined twin morphology.

Note that we have used a nested ADMM to ensure the twinning phase parameter, $\eta$, is resolved before updating the plastic slip and total deformation. Although the twinning sub-algorithm can be included in the main loop, we found better convergence and behavior with this split, in particular when the phase parameter is rapidly evolving.

Finally, to implement the method, we discretize the domain using a uniform grid. The grid must be fine enough that we sufficiently resolve the details of the twin boundary and the deformation within individual grains; i.e. $\frac{L}{\sqrt{N_g}} \gg 8\sqrt{\frac{\alpha}{M}} \gg \frac{L}{N}$ where $L$ is the size of the domain, $N$ is the spatial discretization and $N_g$ is the number of grains.

\section{Pseudo-Slip Model}
We adopt the model of Chang {\it et al.} \cite{Chang2017}. The pseudo-slip model treats the twins averaging over a volume to obtain a twin volume fraction $\lambda$ that is treated similar to plastic slip activity $\gamma$ in crystal plasticity. The deformation gradient is then decomposed multiplicatively as $\mathbf{F} = \mathbf{F}^{\, \text{e}}\mathbf{F}^{\, \text{in}}$ where the inelastic deformation gradient $\mathbf{F}^{\, \text{in}}$ is updated as
\begin{equation}
\dot{\mathbf{F}}^{\, \text{in}}\mathbf{F}^{\, \text{in}-1} = \dot{\gamma^{\, \text{p}}} \Big[ (1 - \lambda) \mathbf{\hat{b}}^{\, \text{p}}\otimes\mathbf{\hat{n}}^{\, \text{p}} + \lambda \mathbf{\hat{b}}^{\, \text{p}'} \otimes\mathbf{\hat{n}}^{\, \text{p}'} \Big] + \dot{\lambda} \gamma^{\, \text{t}} \mathbf{\hat{b}}^{\, \text{t}} \otimes\mathbf{\hat{n}}^{\, \text{t}} \;\;\; \text{w/} \;\;\; \dot{\gamma}^{\, \text{p}} \geq 0.
\end{equation}
Above ($\mathbf{\hat{b}}^{\, \text{p}}, \mathbf{\hat{n}}^{\, \text{p}}$) and ($\mathbf{\hat{b}}^{\, \text{p'}}, \mathbf{\hat{n}}^{\, \text{p'}}$) describe the slip systems in the original and twinned crystals while ($\mathbf{\hat{b}}^{\, \text{t}}, \mathbf{\hat{n}}^{\, \text{t}}$) describe the twin system.

The free energy is decomposed additively,
\begin{equation}
	\mathcal{E}(\mathbf{F}, \dot{\gamma}^{\, \text{p}}, \lambda) = \int_{\Omega}^{} W(\mathbf{F}, \dot{\gamma}^{\, \text{p}}, \lambda) \,d\Omega = \int_{\Omega}^{} \Big(W_{\, \text{e}}(\mathbf{F},\dot{\gamma}^{\, \text{p}}, \lambda) + W_{\, \text{p}}(\gamma^{\, \text{p}}) + W_{\, \text{t}}(\lambda) \Big) \,d\Omega,
\end{equation}
where
\begin{align}
	&W_{\, \text{e}}(\mathbf{F},\dot{\gamma}^{\, \text{p}}) = \frac{1}{2}\mathbf{E}^{\, \text{e}}:\bm{\mathbb{C}}:\mathbf{E}^{\, \text{e}},\\
	&W_{\, \text{p}}(\gamma^{\, \text{p}}) = \sigma^{\, \text{p}}_{\infty} \bigg[ \epsilon^{\, \text{p}} + \frac{\sigma^{\, \text{p}}_{\infty}}{h} \exp\Big( {-h\frac{\epsilon^{\, \text{p}}}{\sigma^{\, \text{p}}_\infty}}\Big) \bigg],\\
	&W_{\, \text{t}}(\lambda) = \frac{1}{2}k\lambda^2.
\end{align}
and $\epsilon^{\, \text{p}} = |\gamma^{\, \text{p}}|$ represents the accumulated plastic strains, $\sigma^{\, \text{p}}_{\infty}$ the ultimate stress of each system, $h$ the hardening of slip, and $k$ is the hardening term for twinning.  The evolution is governed by the dissipation potential 
\begin{equation}
    \Psi^*_{\, \text{p}}\big(\abs{\dot{\gamma}^{\, \text{p}}}\big) = \tau_0^{\, \text{p}}\dot{\gamma}^{\, \text{p}} + \frac{\tau^{\, \text{p}}_0 \dot{\gamma}^{\, \text{p}}_0}{m_{\, \text{p}}+1} \bigg(\frac{\abs{\dot{\gamma}^{\, \text{p}}}}{\dot{\gamma}^{\, \text{p}}_0}\bigg)^{m_{\, \text{p}+1}} \,\,
    \Psi^*_{\, \text{t}}\big(\dot{\lambda} \big) = \tau_0^{\, \text{t}}\dot{\lambda} + \frac{\tau^{\, \text{t}}_0 \dot{\lambda}_0}{m_{\, \text{t}}+1} \bigg(\frac{\dot{\lambda}}{\dot{\lambda}_0}\bigg)^{m_{\, \text{t}}+1}.
\end{equation}
We implement this model using accelerated computational micromechanics as before with the material parameters given in Table \ref{tab:PseudoSlipParameters}.

Note that there are two important differences from the model presented earlier: the first is that we do not have any double-well or gradient term in the twinning, and the second is the inelastic update.  These make this model much simpler to implement.

\begin{table}
\begin{center}
	\begin{tabular}{ |p{2cm}||p{4cm}|p{5cm}|p{3cm}| }
		\hline
		Parameter & Value & Significance & Reference\\
		\hline
		\multicolumn{4}{|c|}{Elastic Parameters} \\
		\hline
		$\lambda_1$ & $25$ $GPa$ & Stiffness $C_{1111}$ term & \cite{Tutcuoglu2019} \\
		$\lambda_2$ & $15$ $GPa$ & Stiffness $C_{1122}$ term & \cite{Tutcuoglu2019} \\
		$\mu$ & 15 $GPa$ & Stiffness $C_{1212}$ term & \cite{Tutcuoglu2019} \\
		\hline
		\multicolumn{4}{|c|}{Twinning Parameters} \\
		\hline
		$\gamma^{\, \text{t}}_0$ & $0.129$ & Shear magnitude  & \cite{Tutcuoglu2019}\\
		$\dot{\lambda}_0$ & $1.0$ $1/s$ & Reference shear rate  & \cite{Tutcuoglu2019} \\
		$m_{\, \text{t}}$ & $1.0$ & Rate hardening  & \cite{Tutcuoglu2019}\\
		$\tau^{\, \text{t}}_0$ & $2$ MPa & Critical resolved shear stress  & \cite{Tutcuoglu2019}\\
		$k$ & $2$ MPa & Twin hardening parameter & \cite{Tutcuoglu2019}\\
		$\theta^{\, \text{t}}$ & $-\frac{\pi}{8}$ & Twin shear angle & --\\
		\hline
		\multicolumn{4}{|c|}{Plasticity Parameters} \\
		\hline
		$m_{\, \text{p}}$ & $0.05$ & Rate hardening & \cite{Chang2015}\\
		$\dot{\gamma}_0^{\, \text{p}}$ & $1.0$ $1/s$ & Reference shear rate & \cite{Chang2015}\\
		$\tau^{\, \text{p}}_0$ & $4$ MPa & Critical resolved shear stress & \cite{Chang2015}\\
		$\sigma^{\, \infty}$ & $2$ MPa & Ultimate slip stress & \cite{Chang2015}\\
		$h$ & $7.1$ GPa & Hardening constant & \cite{Chang2015}\\
		$\theta^{\, \text{p}}$ & $\frac{\pi}{8}$ & Slip shear angle & --\\
		\hline
	\end{tabular}\\
	\caption{Pseudo slip simulation parameters.}
	\label{tab:PseudoSlipParameters}
\end{center}
\end{table}

\newpage

\bibliography{twinslippaper}
\end{document}